\documentclass[review]{elsarticle}

\usepackage{lineno,hyperref}
\modulolinenumbers[5]
\usepackage{amsmath,amssymb}

\journal{Journal of Nuclear Instruments and Methods in Physics Research}

\usepackage{xcolor}

\usepackage{xspace}
\usepackage{slashed}

\newcommand{\ttbar}{\ensuremath{{t\bar{t}}}}
\newcommand{\pt}{\ensuremath{{p_{\rm T}}}}

\newcommand{\Yboost}{y_{\mathrm{boost}}^{\ttbar{}}}
\newcommand{\Chittbar}{\chi^{t\bar{t}}}
\newcommand{\Pout}{p_{\mathrm{out}}}
\newcommand{\mtt}{m^{\ttbar{}}}
\newcommand{\ytt}{y^{\ttbar{}}}
\newcommand{\deltatt}{\delta^{\ttbar{}}}
\newcommand{\abscosthetastar}{|\cos\theta^*|}
\newcommand{\ytlep}{y^{t,\,\mathrm{lep}}}

\newcommand{\Delphes}{\textsc{Delphes}\xspace}
\newcommand{\Rivet}{\textsc{Rivet}\xspace}
\newcommand{\Pythia}{\textsc{Pythia}\xspace}
\newcommand{\GeV}[1]{{#1$\,\textrm{GeV}$}\xspace}
\newcommand{\TeV}[1]{{#1$\,\textrm{TeV}$}\xspace}
\newcommand{\MadGraph}{\textsc{MadGraph}\xspace}

\usepackage{graphicx,epsfig}

\usepackage{url}
\usepackage{hyperref}
\usepackage{ dsfont }

\begin{document}

\begin{frontmatter}

\title{Study of methods of resolved top quark reconstruction in semileptonic $\ttbar$ decay: Erratum}

\author{J. Kvita}
\address{Regional Centre of Advanced Technologies and Materials, Joint Laboratory of Optics of Palack\'{y} University and Institute of Physics AS CR, Faculty of Science, Palack\'{y} University, 17. listopadu 12, 771 46 Olomouc, Czech Republic}


\cortext[mycorrespondingauthor]{J. Kvita}
\ead{jiri.kvita@upol.cz}


\begin{abstract}
Study of methods of resolved top quarks kinematic reconstruction in the $\ttbar \rightarrow \ell+$jets channel is presented at the particle level as well as the fast-simulation detector level. Previous and current pseudo-top quark reconstruction algorithms are compared with suggestions presented on how to improve the reconstructed top-quark mass line shape, including the check of performance on physics observables in terms of correlations between detector, particle and parton levels, and in unfolding, with implications for current high energy physics experiments.
\end{abstract}

\begin{keyword}
HEP\sep pseudo-top quark \sep kinematic reconstruction \sep unfolding
\end{keyword}

\end{frontmatter}


\section{Introduction}

\noindent This paper is an Erratum to~\cite{Kvita:2018trd}.

Top quark is the heaviest fermion in the Standard Model, its large mass~\cite{1674-1137-40-10-100001} leading to a corresponding mean life time below the typical hadronization time, although the decay dynamics is governed by the weak interaction. The top quark decays within the third generation of quarks to a $W$ boson and a $b$ quark in almost 100\% cases.

In hadron collisions, top quarks are produced either singly with the participation of the weak interaction, or in pairs via the strong interaction, although interference between these two leading-order pictures is present in higher orders of the perturbation theory. Production of multiple top quark final states is a subject of experimental searches.

When produced at low transverse momentum ($\pt$) w.r.t. the beam axis ($p^{t}_\mathrm{T} \lesssim m_t / 2$), top quark decay products can be identified via angularly resolved objects in a~detector. With increasing transverse momentum, however, top quark decay products become collimated and merged into ``boosted'' objects requiring dedicated experimental techniques.

While the high-momentum top quarks are interesting in accessing the physics of a~heavy quark at high momentum transfers and possibly probing new physics in the TeV regime, the resolved topology still constitutes the bulk of the statistics delivered in proton-proton ($pp$) collisions by the LHC accelerator and serves as a~useful tool in high energy physics (HEP). Improved methods of top quark identification and reconstruction can thus lead to a~better understanding of not only the physics of the top quark, but also of phenomena where top quark events form a~background to more exotic or beyond-the-standard model (BSM) processes.

Kinematics of resolved top quarks can be reconstructed using the so-called pseudo-top algorithm~\cite{Aad:2015eia} which is a~frequent and useful tool in extracting full kinematic information in the $\ttbar$ environment in $pp$ collisions.
Objects with a~high correspondence to the kinematics of the original top quarks at the parton level are constructed from stable particles or detector-level objects using the same algorithm. Measured and fully corrected (for detector effects) spectra of these objects are used to tune and validate Monte Carlo (MC) generator tunes as well as challenge perturbative quantum chromodynamics (pQCD) calculations at various precision, search for new physics and constrain spectra shapes in the $\ttbar$ sample which is an important background for searches for e.g. the $\ttbar{}+$Higgs boson production.

Measurements unfolded to the particle level in well-defined fiducial phase-space volumes close to the detector level are useful for parameters tuning and validation of fixed-order MC generators at various precision and of different models of processes like hadronization, initial and final state radiation or underlying event~\cite{ATL-PHYS-PUB-2016-020}.

A~solid definition of particle-level objects with a~good correspondence to top-quarks kinematics is important in order not to dilute the information at both detector and particle levels. Using parton-level top quarks as the reference level to which measured spectra are corrected involves large corrections to the full phase-space as well as theoretical ambiguities of defining top quarks as partons.
The definition of variables at the particle level with a~good correlation to the four-momenta of parton top quarks is preferred as it provides a~weaker model dependence of the measured cross sections compared to the definition at the parton level, yielding more robust results in time as a~heritage of current high-energy physics experiments.

The goal of the presented study is to compare various modifications of the pseudo-top algorithm and their performance in terms of the resolution of the reconstructed top quark and $\ttbar{}$ mass as well as in terms of the degree of correlation between parton, particle and detector levels.
The physics objects and event selection are described in~Section~\ref{sec:select}.
Events where $\ttbar{}$ pairs are produced in $pp$ collisions at the central-mass-energy of \TeV{13} were generated at particle level, with the subsequent detector level simulated using simple yet realistic tools as described in~Section~\ref{sec:samples} with the focus on the approximate ATLAS experiment geometry and resolutions. Only events in the semileptonic $\ttbar$ decay channel are generated as this channel provides optimal signal-to-background ratio and large statistics in current experimental data, as well as reasonably constraint kinematics. Results are presented in~Section~\ref{sec:results} while \ref{app1} summarizes the analytic solutions to various conditions used to reconstruct the missing kinematic information carried away by the neutrino.

\section{Objects Definition and Selection}
\label{sec:select}

This study focuses on cases where the $\ttbar{}$ pair decays semileptonically, i.e. one $W$ boson from either top quark decays hadronically while the other decays leptonically into a pair of a lepton and a neutrino. Decays to a $\tau$ lepton are considered when the $\tau$ lepton decays to an electron or a muon (and the corresponding neutrino), which can then pass the selection criteria.

The \Rivet{}~\cite{Buckley:2010ar} version 3.0.2 and the \Rivet{} analysis \texttt{ATLAS\_2015\_I1404878}~\cite{Aad:2015mbv} of the 8 TeV measurement of differential spectra in $pp \rightarrow \ttbar{}$ events by the ATLAS experiment have been used as the baseline of the objects selection and the pseudo-top algorithm definition, which was then modified (see Section~\ref{sec:results}).

Collimated hadronic final states dubbed ``jets'' reconstructed from stable particles except neutrinos by the Anti-$k_t$ algorithm~\cite{Cacciari:2008gp} with the distance parameter of 0.4 are required to be within pseudorapidity~\footnote{The pseudorapidity $\eta$ is defined using the polar angle $\theta$ from the positive $z$ axis coinciding with one of the colliding proton beam as $\eta \equiv -\ln\tan\frac{\theta}{2}$.} $|\eta| < 2.5$ and to pass the requirement on their transverse momentum (w.r.t. the beam, i.e. the $z$, axis) of \GeV{$\pt > 25$}. Jets are further labelled (tagged) as $b$-jets if a $b$-hadron with $\pt > 5\,$GeV is found within $\Delta R < 0.4$ around the jet axis. The presence of two $b$-jets is an important event signature and is part of most event selection in HEP analyses concerning top quarks.

Leptons (electron or muons) are selected within the same kinematic limits, but are first ``dressed'' in terms of adding four-momenta of photons within 0.1 in a cone of radius defined as $\Delta R = \sqrt{\Delta\eta^2 + \Delta\phi^2}$ around the lepton, to account for final-state photon radiation which typically is included in the lepton final states in a~detector. Particle jets overlapping with the selected lepton within $\Delta R < 0.2$ are removed.

In summary, at least four jets are expected in the event, two of which are required to be $b$-tagged, and a high-$\pt$ lepton and a large transverse energy imbalance in the event due to the escaping neutrino. In practice, the requirement of two $b$-jets often yields sufficiently pure $\ttbar{}$ sample that additional selection criteria on the missing transverse energy are not needed.
While events with one $b$-tagged jets are often used e.g. for measuring the inclusive cross-section, they are not considered in this study as the requirement of two $b$-tagged jets removes combinatorial ambiguities in the jet assignment to top quark decay products.

\section{Samples}
\label{sec:samples}

All events were generated for the case of $pp$ collisions at the centre-of-mass energy of \TeV{13} using the \MadGraph{} version {\tt 2.5.5} simulation toolkit~\cite{Alwall:2014hca} which was chosen for its versatility and ability to generate all processes considered in this analysis. This generator has also been used for data comparison by the CMS collaboration and gradually also by the ATLAS collaboration.
In total, $2\,$M events were generated for parton-shower-to-matrix-element matched processes $pp \rightarrow \ttbar{}+$jet at the leading (LO) order in pQCD and $pp \rightarrow \ttbar{}$ at the next-to-leading (NLO) order using the Standard Model matrix elements. For the purpose of studying the unfolding performance, an alternative sample of 2~M $\ttbar{}$ events was generated at the LO only, to provide a sample with slightly different spectra.
Finally, 1~M events were generated for the process of a hypothetical additional neutral heavy vector boson $Z'$ decaying as $pp \rightarrow Z' \rightarrow \ttbar{}$ (using the model \cite{FeynModelZprime,Christensen:2008py,Wells:2008xg}). Parton shower and hadronization were simulated using the integrated \Pythia{}8~\cite{Sjostrand:2007gs,Pythia8} generator and the top-quark mass of $172\,$GeV (\MadGraph{} default) was used for all simulated samples. The detector-level simulation is described in~Sect.~\ref{sec:delphes}.

\section{Pseudotop algorithm studies}
\label{sec:results}

\subsection{Hadronic pseudo-$W$}

\begin{figure}[!h]
  \includegraphics[width=0.50\textwidth]{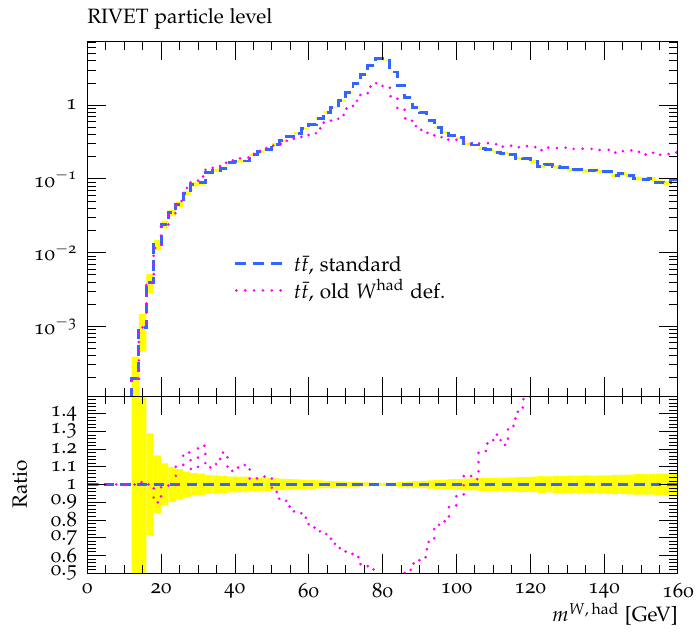}
  \includegraphics[width=0.50\textwidth]{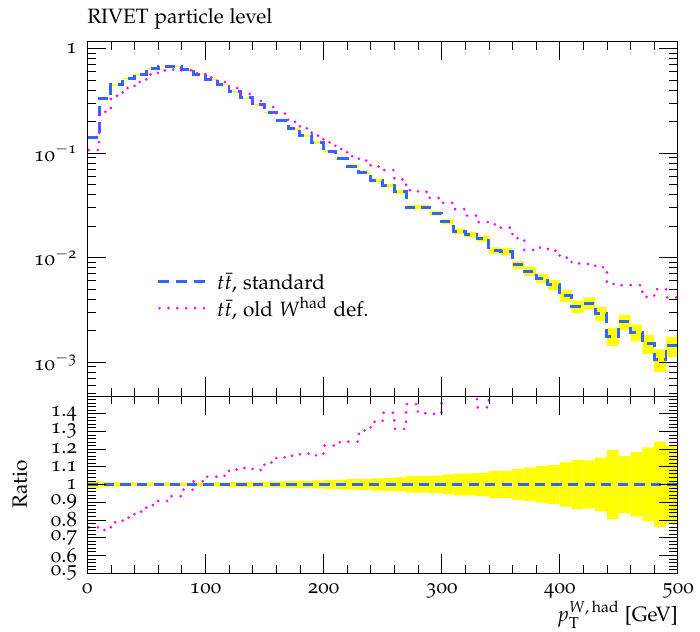}
  \\
  \includegraphics[width=0.50\textwidth]{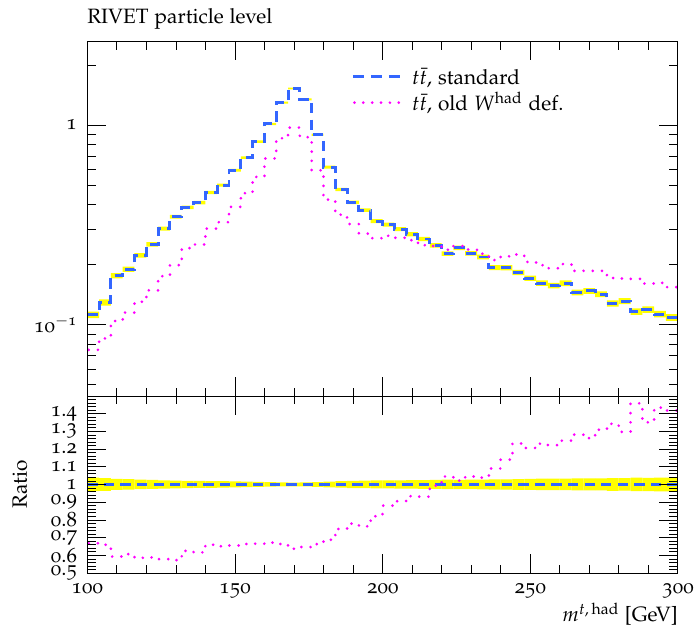}
  \includegraphics[width=0.50\textwidth]{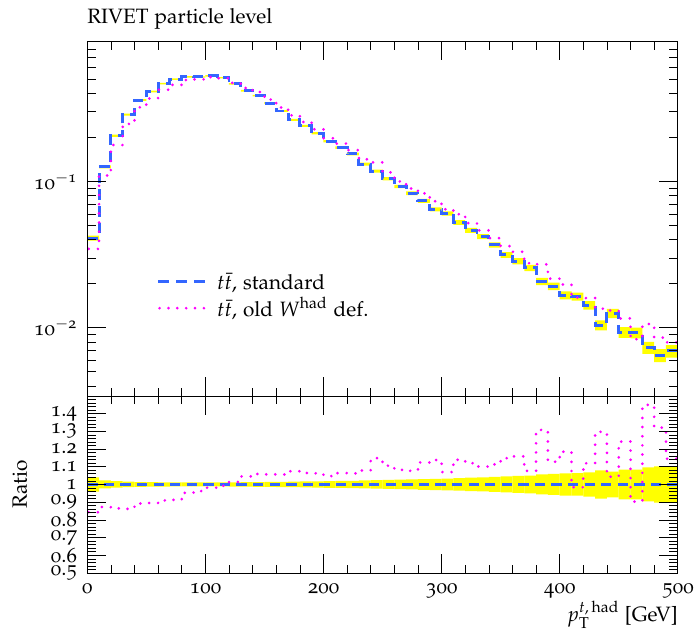}
  \caption{The particle-level hadronic pseudo-$W$ mass (top left) and \pt{} (top right), and the hadronic pseudo-top quark mass (bottom left) and \pt{} (bottom right) for different choices of the light jets to form the hadronic pseudo-$W$ in the event: as the pair of jets with invariant mass closest to $m_W$ (dashed), or as the pair of highest-\pt{} non-$b$-tagged jets (dotted).
  Ratios to the standard option are provided in lower panels, the yellow band indicating the statistical uncertainty in the denominator.}
\label{pst:mttlep_nupz_study1}
\end{figure}

The particle-level candidate for the hadronically decaying $W$ boson is composed from non-$b$-tagged jets by either using such two highest-\pt{} light jets or by finding the pair of light-flavour jets with an invariant mass closest to the $W$ boson mass $m_W = 80.4\,$GeV. The two scenarios, as defined and used in ATLAS 7~TeV~\cite{Aad:2015eia}; and ATLAS 8~TeV~\cite{Aad:2015mbv} and 13~TeV~\cite{Aaboud:2017fha} analyses, respectively, are compared at the particle level in Fig.~\ref{pst:mttlep_nupz_study1} using the privately simulated samples as detailed in~Sec.~\ref{sec:samples}.
The plots show that the original definition (denoted ``old $W^\mathrm{had}$'' in plot legends) using the pair of highest-\pt{} non-$b$-tagged jets was improved (in what is now the standard option) by using the pair of jets with invariant mass closest to $m_W$. Improvement is seen terms of the line shapes of both the hadronic pseudo-$W$ and hadronic pseudo-top masses ($m^{W,\mathrm{had}}$ and $m^{t,\mathrm{had}}$), namely providing a~less-pronounced tail towards larger masses. A~change in slope of the transverse momentum spectra ($p_{\mathrm{T}}^{W,\mathrm{had}}$ and $p_{\mathrm{T}}^{t,\mathrm{had}}$) is also seen, although if reproduced at both particle and detector levels this is not a~priory a~problem in using either definition e.g. for MC tuning studies. Still, any improvement in the mass line shape is of course a~preferred option, as it possibly improves also the correlation to the parton level.

\subsection{Pseudo-top quarks}

The four-momentum of the leptonically decaying pseudo-top quark is defined by adding the four-momenta of the $b$-jet closest to the lepton and of the reconstructed leptonically decaying pseudo-$W$ candidate detailed later. Finally, the four-momenta of the hadronically decaying pseudo-top quark is defined as the sum of the four-momenta of the remaining highest-\pt{} $b$-jet and of the hadronic pseudo-$W$ candidate.

\subsection{Optimization of the $p_z^\nu$ choice}

As the undetected neutrino from the leptonic $W$ decay carries away kinematic information, its momentum has to be reconstructed. The transverse component of its momentum can be easily estimated using the vector of the reconstructed missing transverse energy, defined as the negative sum of the neutrinos transverse momenta at the particle level or as the negative sum of calorimeter transverse energy deposits at the detector level. Neutrino's longitudinal momentum ($p_z^\nu$) has to be computed from an additional reasonable physics constrain. The following choices are tried for the computation of $p_z^\nu$ and compared for distributions of rapidities of the leptonic $W$ and leptonic top quark ($y^{W,\mathrm{lep}}$ and $y^{t,\mathrm{lep}}$) and checking also their hadronic counterparts ($y^{W,\mathrm{had}}$ and $y^{t,\mathrm{had}}$).

\begin{figure}[p]
  \includegraphics[width=0.50\textwidth]{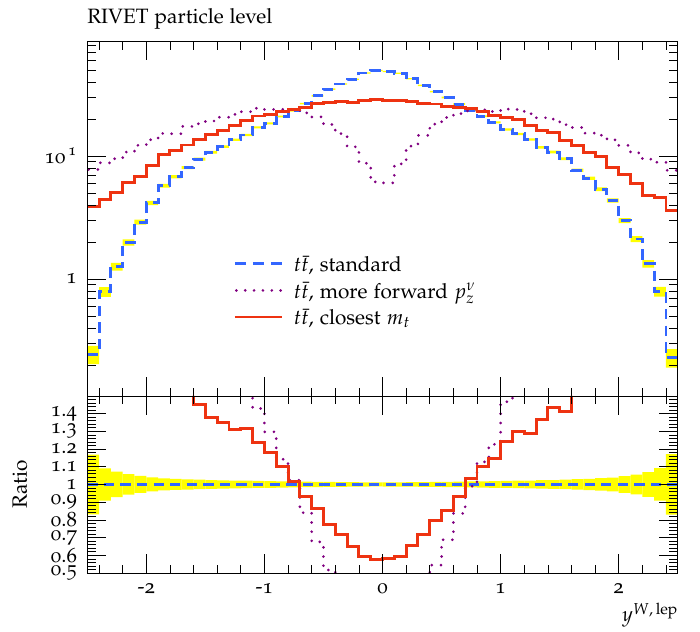}
  \includegraphics[width=0.50\textwidth]{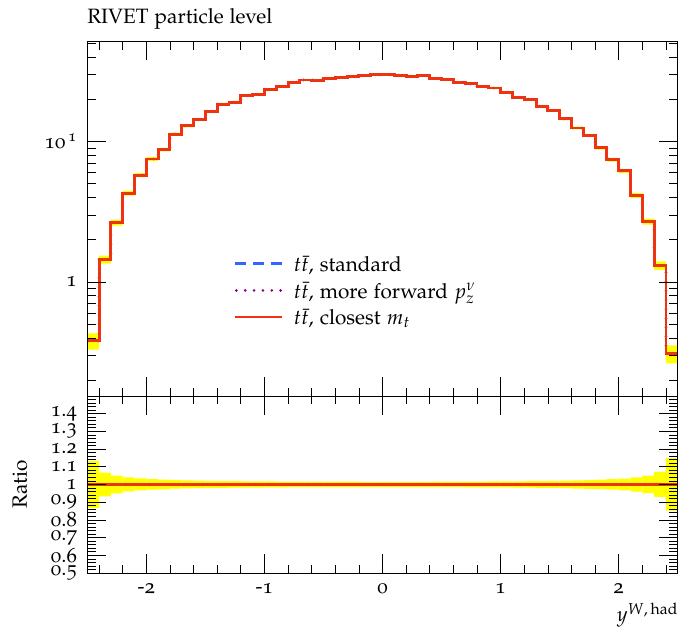}
  \\
  \includegraphics[width=0.50\textwidth]{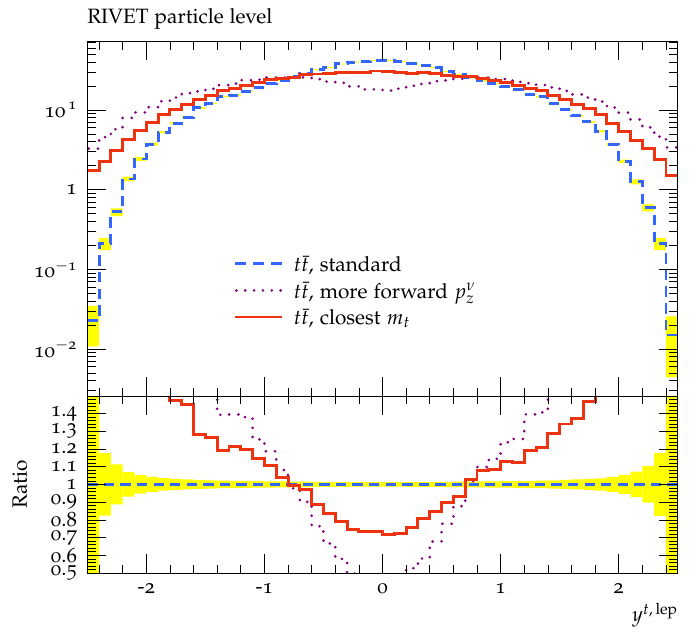}
  \includegraphics[width=0.50\textwidth]{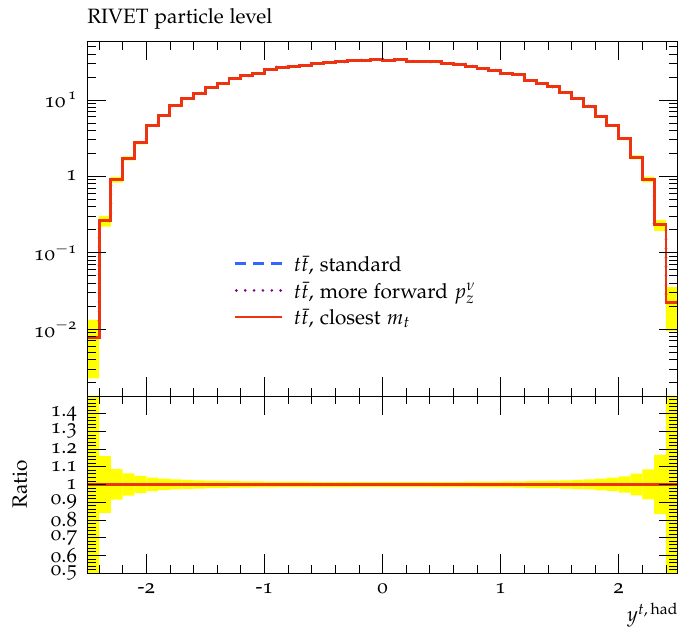}
  \caption{The particle-level leptonic (top left) and  hadronic (top right) pseudo-$W$ rapidity and leptonic (bottom left) and hadronic (bottom right) pseudo-top rapidity for different choices of the neutrino $p_z$ solution based on the $m_{\ell\nu}= m_W$ condition: the standard choice (dashed), 
    more forward (dotted) and 
    ``closest $m_t$'' (solid).
      Ratios to the standard option are provided in lower panels, the yellow band indicating the statistical uncertainty in the denominator.
  }
\label{pst:Wrap_cmp2}
\end{figure}

\begin{figure}[p]
  \includegraphics[width=0.50\textwidth]{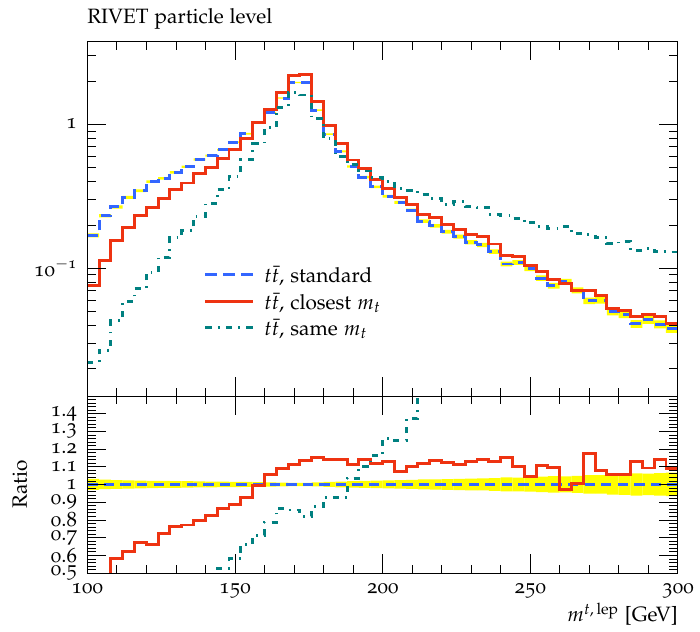}
  \includegraphics[width=0.50\textwidth]{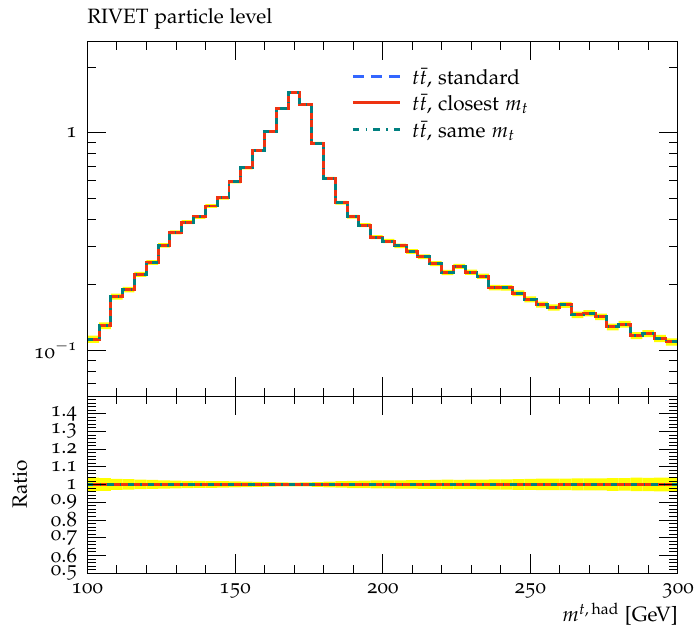}
\\
  \includegraphics[width=0.50\textwidth]{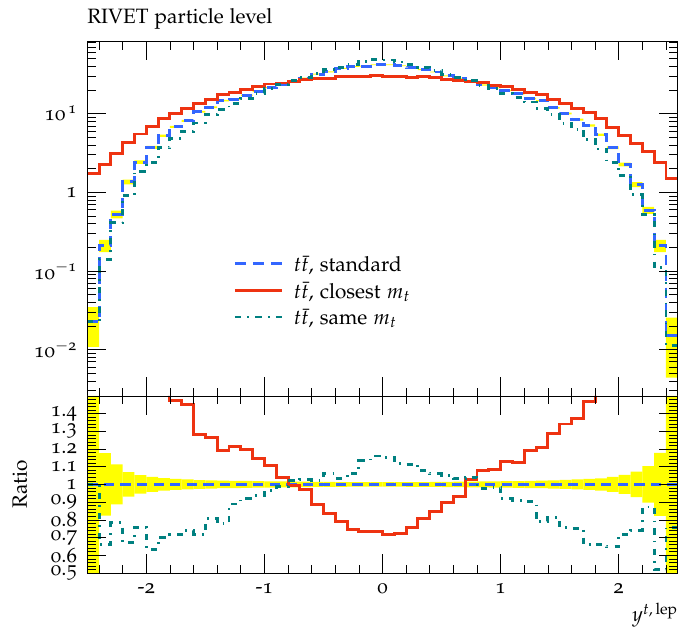}
  \includegraphics[width=0.50\textwidth]{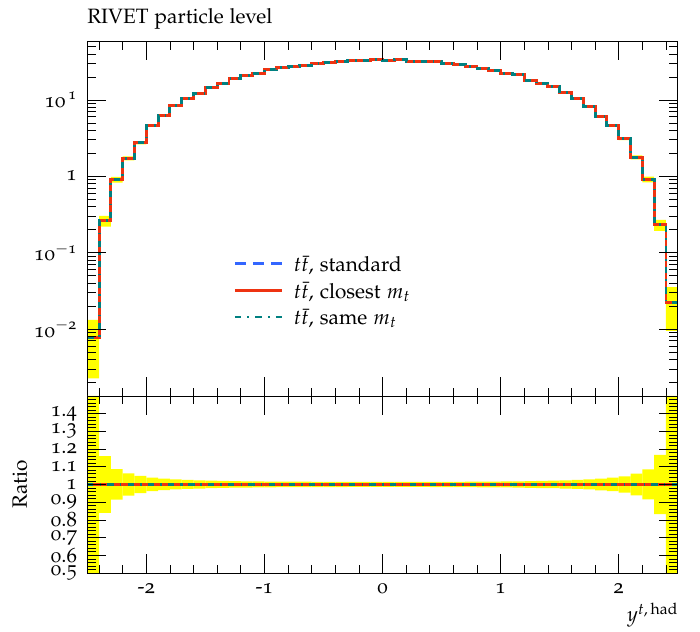}
  \caption{The particle-level leptonic (top left) and hadronic (top right) pseudo-top mass and rapidity of the leptonic (bottom left) and hadronic (bottom right) pseudo-top for different choices of the neutrino $p_z$ solution: the standard choice (dashed),
    ``closest $m_t$'' (solid); and 
    ``same $m_t$'' (dot-dashed). The hadronic pseudo-top spectra are unaffected by the choices on the leptonic side of the event, showing however the similarity of the $y_{t,\mathrm{had}}$ rapidity spectrum to the leptonic one from the ``closest $m_t$'' solution.
    Ratios to the standard option are provided in lower panels, the yellow band indicating the statistical uncertainty in the denominator.
  }
\label{pst:pseudotop_nupz_mt_study3}
\end{figure}

\begin{enumerate}

  \item The usual (denoted as ``standard'' in plot legends) definition of the leptonic pseudo-$W$ and leptonic pseudo-top relies on the solution of $p^\nu_z$ from a~quadratic equation stemming from the $m_{\ell\nu}= m_W$ condition.
If a~complex solution is found, the imaginary part is dropped, when two real solutions exist, the one with smaller $|p_z^\nu|$ is taken.
This choice has some physics motivation, e.g. in the fact that top quark pairs are produced in $gg$, i.e. same-parton species, collisions, and on average a~large imbalance in the $p_z$ of the $gg$ system is not expected. However, this neutrino solution leads to visibly different spectra of rapidities of leptonic pseudo-$W$ and pseudo-top quark candidates (see Fig.~\ref{pst:Wrap_cmp2}), compared to those of their hadronic counterparts, namely being significantly more central by construction.

  \item As a~test and a~check, the more forward  $p_z$ solution is also tried, denoted as ``more forward'' in plot legends.

  \item As a~modification, a~new condition (denoted as ``closest $m_t$'') based on the minimal difference $|m_{t,\mathrm{had}} - m_{t,\mathrm{lep}}|$ is used to choose the best $p_z^\nu$ solution.
This simple reconsideration leads to a~rapidity spectrum of the leptonic pseudo-$W$ as well as of the leptonic pseudo-top be closer in shape to those of their hadronic counterparts (see Fig.~\ref{pst:Wrap_cmp2}--\ref{pst:pseudotop_nupz_mt_study3}), though slightly broader.
However, as seen in Figure~\ref{pst:pseudotop_nupz_mt_study3}, the leptonic pseudo-top mass spectrum is improved in the low-mass tail and especially in the peak of the distribution.

  \item Next, a~novel solution (denoted as ``same $m_t$'' in plot legends) to the $p^\nu_z$ problem is defined as a~solution to the  $m_{t,\mathrm{had}} = m_{t,\mathrm{lep}}$ condition, taking again a~more central solution  in case of a~positive quadratic equation discriminant (see again Fig.~\ref{pst:pseudotop_nupz_mt_study3}). Although this algorithm further diminishes the low-mass tail for the leptonic pseudo-top, it largely increases the large-mass tail and decreases magnitude in the peak region and leads to large tails in the mass distribution of the leptonic pseudo-$W$ (not shown).

  \item Returning to the $p_z^\nu$ solution from the $m_{\ell\nu}= m_W$ condition, a~swap in the $b$-jets assignment is also newly allowed, and both neutrino solutions are also tried similarly as in the ``closest $m_t$'' solution, so in total the best choice out of four is selected in terms of minimal $|m_{t,\mathrm{had}} - m_{t,\mathrm{lep}}|$; this algorithm is denoted ``best $m_t$'' in plot legends.

\end{enumerate}
Other methods, like trying the ``same $m_t$'' solution first when in the case of a~negative discriminant the standard solution is tried next, were also tested, but these approaches did not lead to significant improvements in performance.

\subsection{Performance on the line shape of a~hypothetical $Z'$ particle}

\begin{figure}[!h]
  \includegraphics[width=0.50\textwidth]{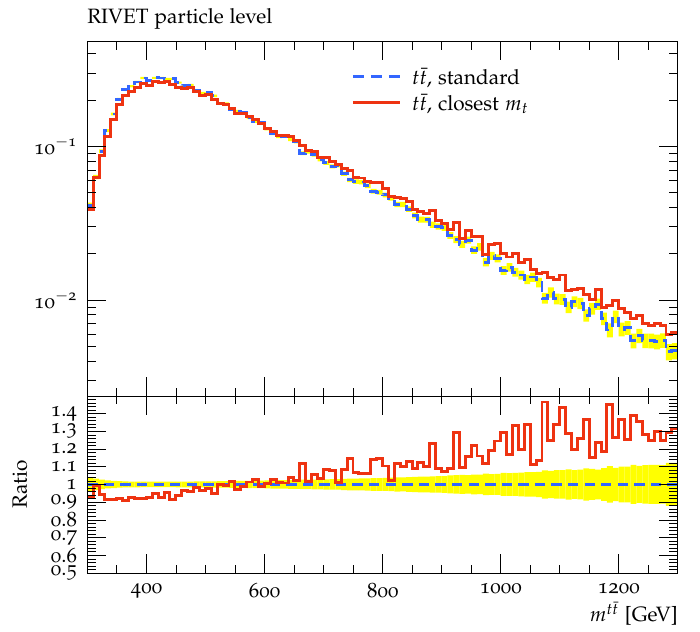}
  \includegraphics[width=0.50\textwidth]{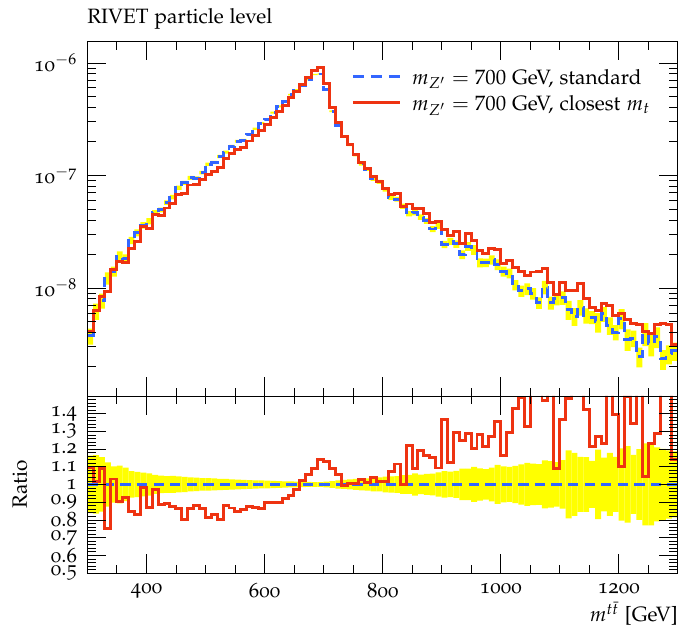}
  \caption{The particle-level pseudo-\ttbar{} invariant mass distribution for the \ttbar{} sample (left) and for the hypothetical $Z'$ boson of mass of $700\,$GeV and decaying to a $\ttbar$ pair for the different choices of the neutrino $p_z$ solution based on the $m_{\ell\nu}= m_W$ condition: the standard choice (dashed) 
    and ``closest $m_t$'' (solid).
      Ratios to the standard option are provided in lower panels, the yellow band indicating the statistical uncertainty in the denominator.
  }
\label{pst:mtt_nupz_study4}
\end{figure}

Performance of one of the new choice of the neutrino $p_z$ solution w.r.t. the standard one was checked on the shape of the reconstructed mass peak of a hypothetical particle $Z'$ particle decaying to a $\ttbar$ pair. Its mass of $m_{Z'} = 700\,$GeV was selected such that the resolved topology of top quark decay products is still dominant over the boosted one. The results are presented in~Fig.~\ref{pst:mtt_nupz_study4}, showing a~sharper peak of the pseudo-\ttbar{} mass ($\mtt$) distribution for the novel proposed method (``closest $m_t$'').

Of course, the physical binning is driven by the experimental resolution and cannot be this fine, however, a~$10$\% improvement in the peak region is possible, which is at the level of the typical experimental uncertainties and resolution.

\subsection{Kinematic variables}
\label{sec:vars}
By construction, the changes in the $p_z^\nu$ choice do not affect $\pt$-related quantities of the leptonic top quark nor the \ttbar{} system, nor the out-of-plane variable $p_{\rm out}$~\cite{Aad:2015mbv} used in initial and final state radiation tuning~\cite{ATL-PHYS-PUB-2016-020}.
However, improvement may be searched for in the line shape of the mass and rapidity of the leptonic top quark ($m^{t,\mathrm{lep}}$, $y^{t,\mathrm{lep}}$) and of the \ttbar{} system ($\ytt$), and the mass ($\mtt$) of the \ttbar{} system, or other variables composed from the two top quarks which also use the longitudinal momentum, like the $\cos\theta^*$ (angle between a~top quark and the $z$ axis in a~frame where the $\ttbar{}$ system has zero momentum along the $z$ axis) and the laboratory opening angle between the two top quarks ($\delta_{\ttbar{}}$).
Further variables studied later are the transverse momentum of the $\ttbar$ system ($p_\mathrm{T}^{\ttbar{}}$) and the out-of-plane momentum $p_{\mathrm{out}}$ which has two entries per event due to the possible r\^{o}le swap of a~top quark to define a~plane together with the $z$ axis direction, to which the momentum of the other top quark is projected; and the $\Yboost$ and $\Chittbar$ variables, defined as
$$ \Pout \equiv  \vec{p}^{\,t, \mathrm{had}} \cdot \frac{\vec{p}^{\,t,\mathrm{lep}} \times \hat{z}}{|\vec{p}^{\,t,\mathrm{lep}}\times \hat{z}|}   \,, \quad \mathrm{and\,\, had \leftrightarrow lep} $$
$$ \Yboost \equiv \frac12 \left| y^{t,\mathrm{had}} + y^{t,\mathrm{lep}} \right| $$
$$ \Chittbar \equiv \exp \left| y^{t,\mathrm{had}} - y^{t,\mathrm{lep}} \right| \,. $$
These are sensitive to final state radiation, the boost of the $\ttbar{}$ system and thus also to PDFs; and to new physics via their sensitivity to the production angle in central mass system. Their shapes also differ for the ``same $m_t$'' and ``best $m_t$'' options.

\subsection{Performance on the \Delphes{} detector level}
\label{sec:delphes}
In order to check a~possible improvement in the correspondence between particle and detector levels, the \Delphes{} simulation package~\cite{deFavereau:2013fsa} was used with a~modified ATLAS card (to allow storage of partons, $b$-hadrons and photons needed for dressing of leptons) to simulate the passage of particles through a~realistic particle detector.
The ATLAS card was validated by \Delphes{} authors as described in Section~5 of~\cite{deFavereau:2013fsa}. A cross-check of using similarly modified CMS card in this analysis was also performed, finding very similar results. 
For these studies, $2\,$M $\ttbar{}$ events were generated by \MadGraph{} to provide a~larger sample also at the \Delphes{} detector level due to finite detector efficiency to select the objects within the phase-space defined in~Section~\ref{sec:select}. The efficiency was found to be about 7\%, similar as in real experiments and analyses.

Independent implementations of the aforementioned pseudo-top algorithms were used both at the particle (using \Pythia{}8 stable particles) and \Delphes{} detector levels. 

At the particle level, selected leptons (electrons or muons) were dressed by photons with $\Delta R < 0.1$ w.r.t. the lepton.
The $b$-tagging at the particle level was performed by matching a~particle jet to an open-beauty $b$-hadron (meson or a~baryon) with $\pt > 5\,$GeV based on the PDG-ID codes~\cite{1674-1137-40-10-100001}. If a~match was found within $\Delta R < 0.4$, the particle jet was considered as $b$-tagged.

\begin{figure}[p]
  \includegraphics[width=0.99\textwidth]{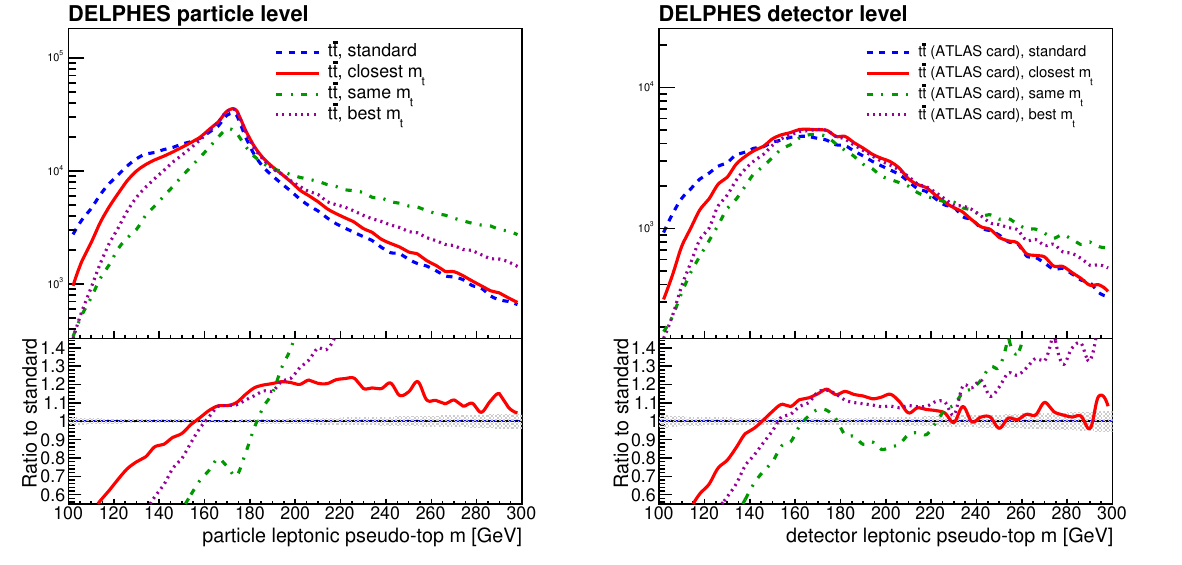}
  \\
  \includegraphics[width=0.99\textwidth]{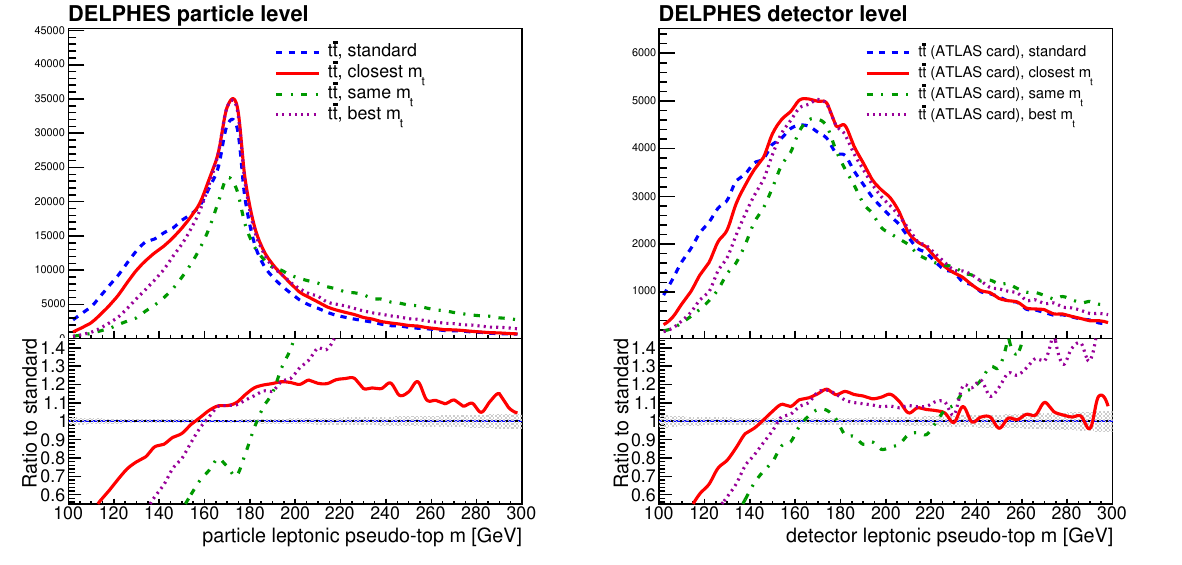}
  \caption{Leptonic pseudo-top mass
    for different choices of the neutrino $p_z$ solution: the standard choice (dashed), 
    ``closet $m_t$'' (solid), 
    ``same $m_t$'' (dot-dashed), and the one giving the best top quark masses allowing also the $b$-jets swap (``best $m_t$'', dotted). 
    Left: particle level, right: \Delphes{} detector level obtained using the ATLAS card.
    Ratios to the standard option are provided in lower panels, the yellow band indicating the statistical uncertainty in the denominator.
    Top (bottom) plots are in the logarithmic (linear) scale.
}
\label{pst:mt_nupz_study5}
\end{figure}

First the performance on the line shape of the leptonic pseudo-top mass is checked in~Fig.~\ref{pst:mt_nupz_study5}, showing a~very similar behaviour compared to the pure \Rivet{} study in the preceding Section at the particle level, and a~slightly modified performance at the \Delphes{} detector level where the new approach (``closest $m_t$'') still yields smaller low-mass tail while the ``same $m_t$'' yields a~slightly sharper peak, although producing a~more pronounced tail to higher masses.
The ``best $m_t$'' choice yields even smaller low-mass tail, but returns even more pronounced tail towards larger masses.

\begin{figure}[p]
  \includegraphics[width=0.99\textwidth]{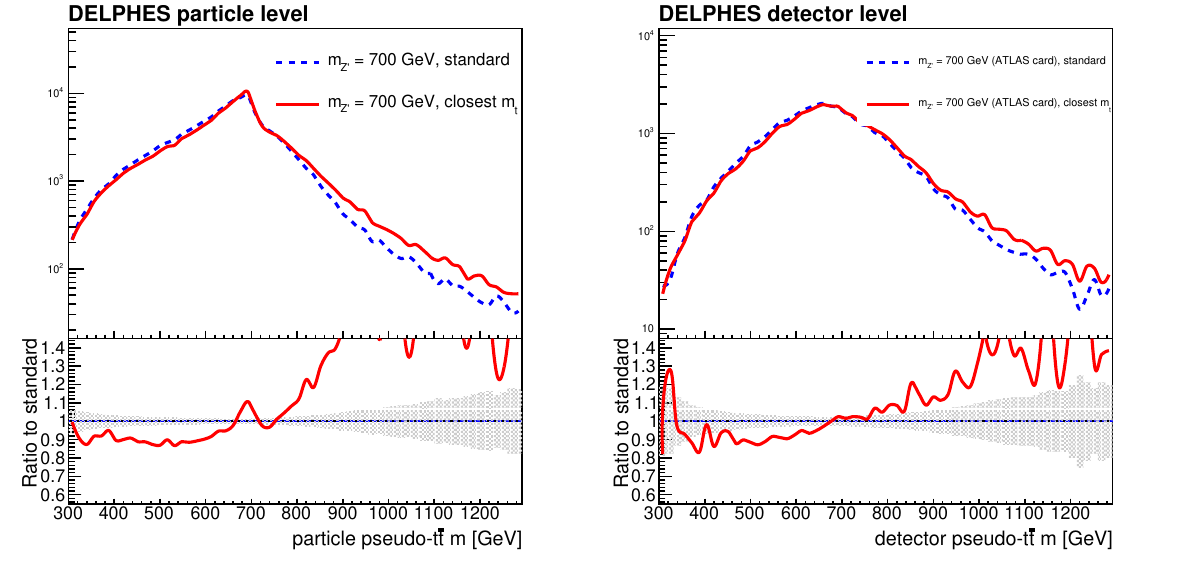}
  \includegraphics[width=0.99\textwidth]{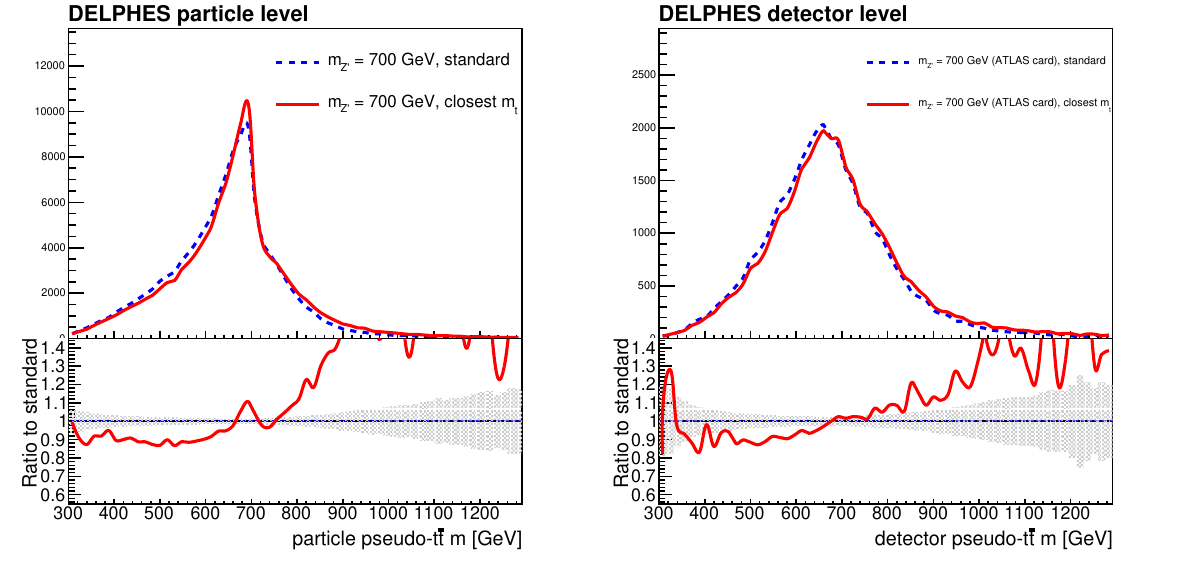}
  \caption{Pseudo-\ttbar{} invariant mass distribution for the hypothetical $Z'$ boson generated at mass of $700\,$GeV and decaying to a $\ttbar$ pair at the particle level (left) and \Delphes{} detector level obtained using the ATLAS \Delphes{} card (right) for the different choices of the neutrino $p_z$ solution based on the $m_{\ell\nu}= m_W$ condition: the standard choice (dashed)
    and the ``closest $m_t$'' (solid).
    Ratios to the standard option are provided in lower panels, the yellow band indicating the statistical uncertainty in the denominator.
    Top (bottom) plots are in the logarithmic (linear) scale.
  }
\label{pst:zp_study8}
\end{figure}

The performance on the hypothetical $Z'$ particle (using the sample of $1\,$M events) at the \Delphes{} detector level is compared in Fig.~\ref{pst:zp_study8} showing unfortunately a~completely washed-out peak compared to a~more pronounced peak of the ``closest $m_t$'' at the particle level, similar to what was found using \Rivet{} in the previous Section.

\subsection{Correlations between levels}
\label{sec:migra}

\begin{figure}[p]
  \includegraphics[width=1.00\textwidth]{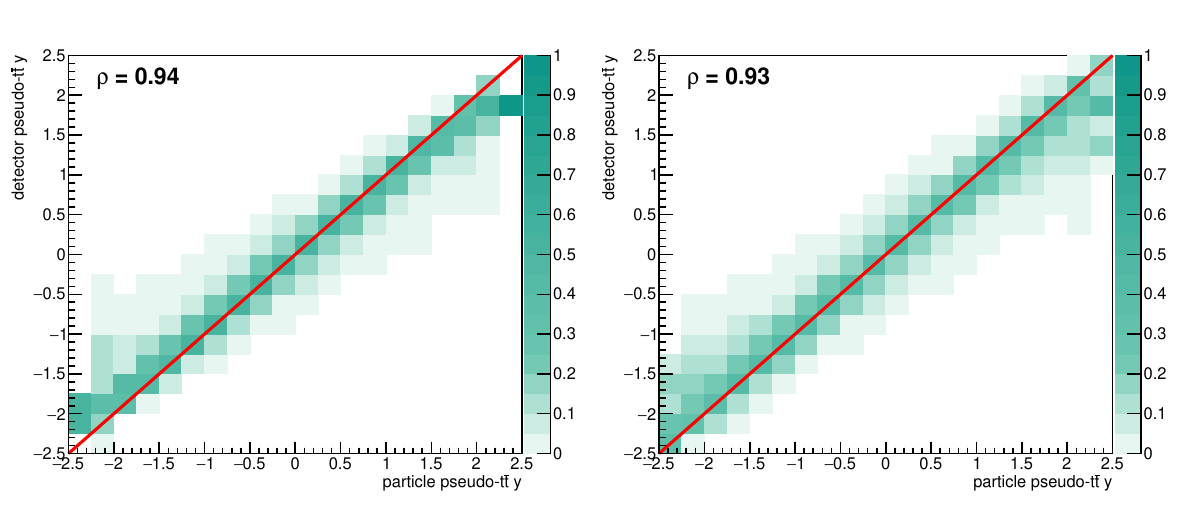}
  \\
  \includegraphics[width=1.00\textwidth]{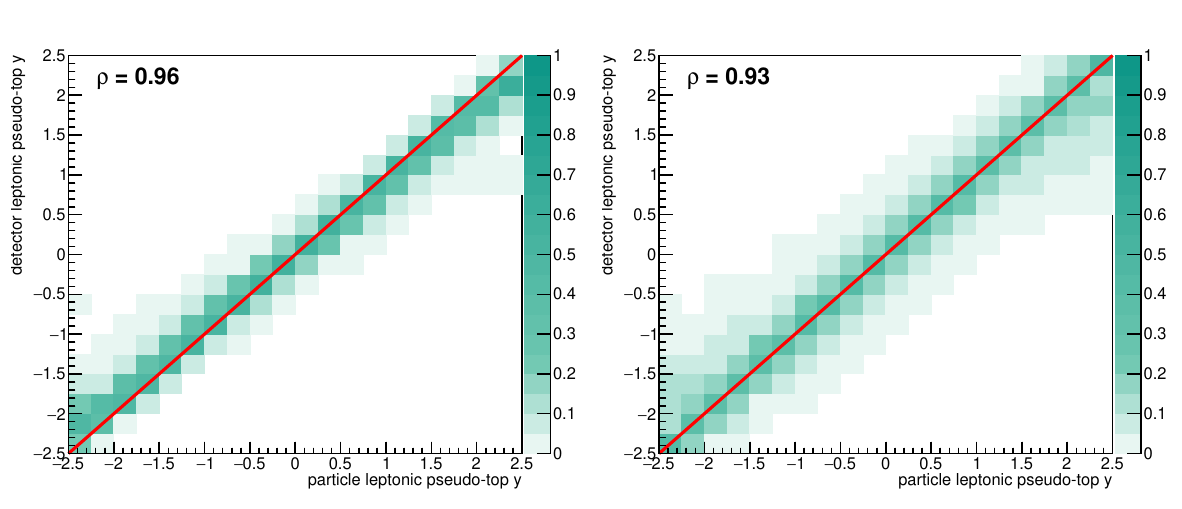}
    \caption{Migration matrices between the particle and the \Delphes{} detector levels for the pseudo-\ttbar{} 
      rapidity (top) 
      and the leptonic pseudo-top rapidity (bottom) 
      obtained using the ATLAS card; for different choices of the neutrino $p_z$ solution based on the $m_{\ell\nu}= m_W$ condition: the standard choice (left) 
      and the ``closest $m_t$'' (right). 
      No angular matching between the particle and detector level objects forming the pseudo-tops was performed.
      The solid bold line is the diagonal, $\rho$ stands for the correlation coefficient evaluated before the normalization of columns.  }
\label{pst:migra_study6_ptcl_det}
\end{figure}

Migration matrices between the particle and the detector (provided by \Delphes{}) levels were obtained and normalized so that each element of the matrix $\mathcal{M}_{ij}$ stands for the fraction of events migrating from a~given particle-level bin $i$ to various detector-level bins labelled $j$.
As rapidities of the leptonic pseudo-top quark and of the $\ttbar$ system depend on the choice of the neutrino $p_z$ solution, migration matrices for these variables were studied. Compared to the standard choice, worse performance in terms of the correlation between the particle and detector levels was found for the ``same $m_t$''  method (not shown) while similar (though slightly lower) for the ``closest $m_t$'' method, as displayed in~Fig.~\ref{pst:migra_study6_ptcl_det}.
Correlations between the particle and detector levels for more kinematic variables and all the studied algorithms are summarized in Tab.~\ref{tab:corrs:particle_detector}.

\subsection{Matching between the particle and detector levels}
\label{sec:match}

\begin{figure}[p]
  \includegraphics[width=1.00\textwidth]{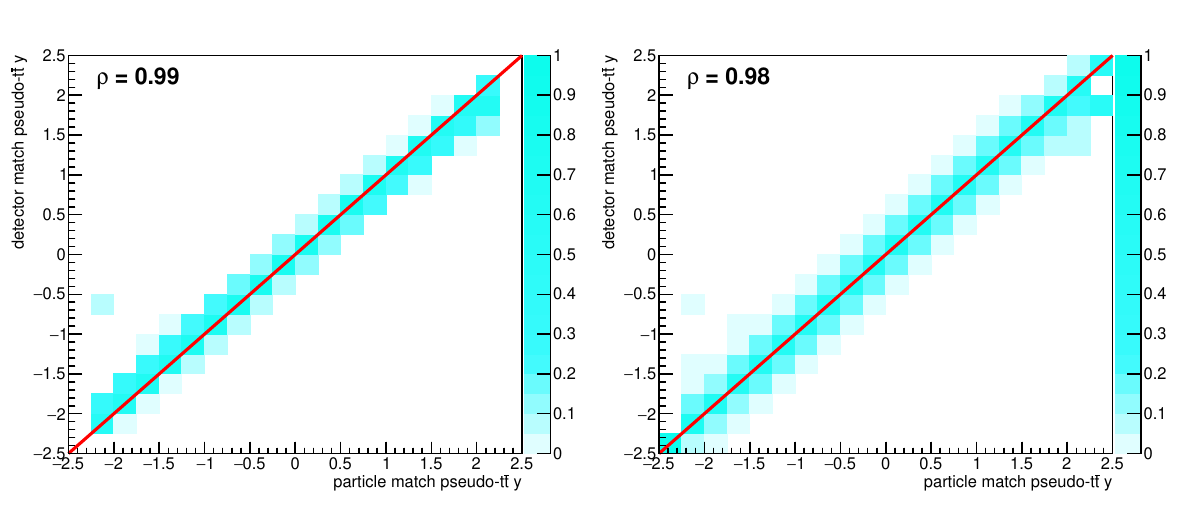}
  \\
  \includegraphics[width=1.00\textwidth]{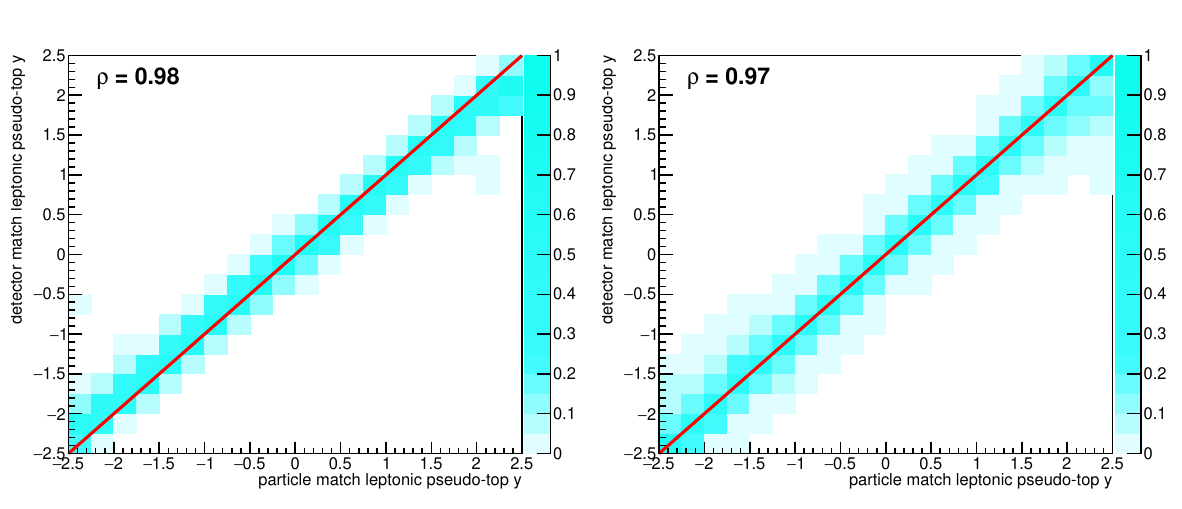}
  \caption{Migration matrices between the particle and the \Delphes{} detector levels for matched events for the pseudo-\ttbar{} 
    rapidity (top) 
    and the leptonic pseudo-top rapidity (bottom) 
    obtained using the ATLAS card; for different choices of the neutrino $p_z$ solution based on the $m_{\ell\nu}= m_W$ condition: the standard choice (left)
     and the ``closest $m_t$'' (right). 
    The solid bold line is the diagonal, $\rho$ stands for the correlation coefficient evaluated before the normalization of columns.  }
\label{pst:migra_study6_ptcl_det_match}
\end{figure}

\begin{figure}[p]
  \includegraphics[width=1.00\textwidth]{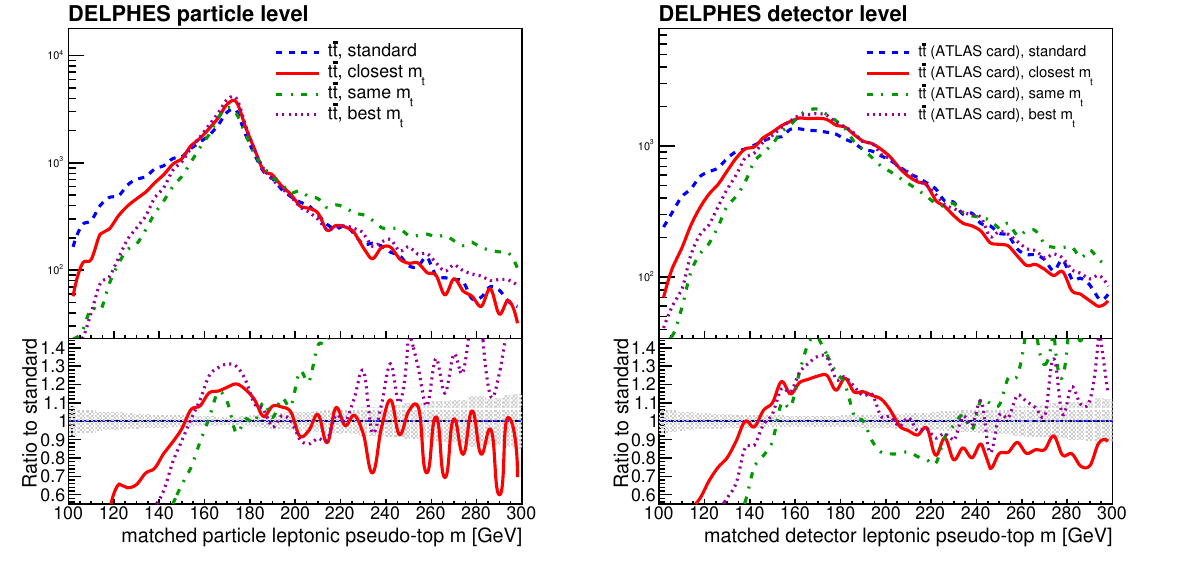}
\\
  \includegraphics[width=1.00\textwidth]{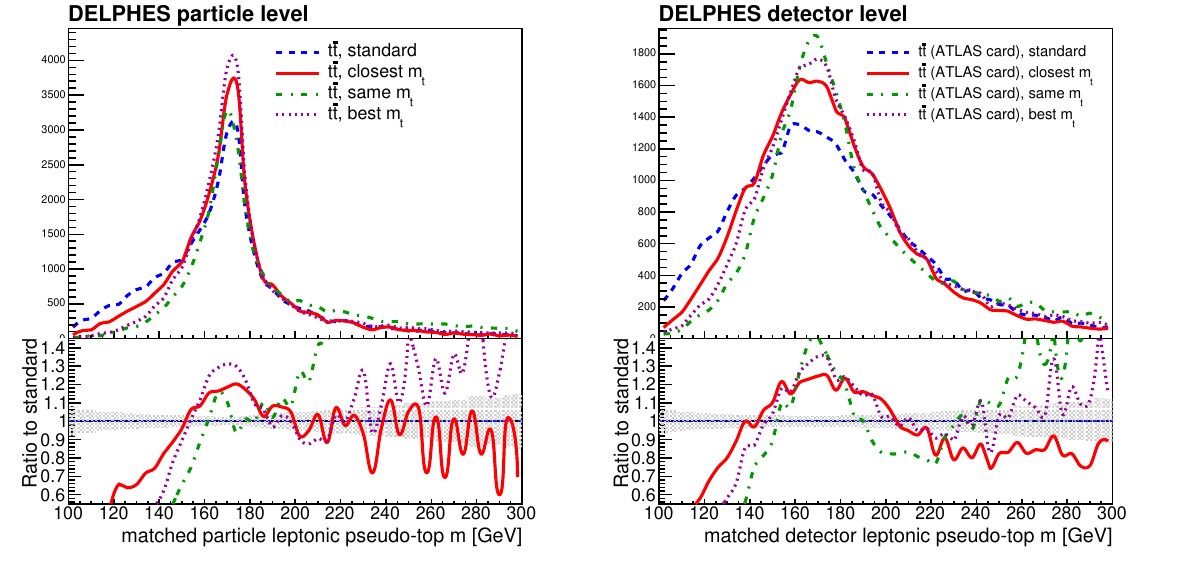}
  \caption{Leptonic pseudo-top mass for matched events 
    for different choices of the neutrino $p_z$ solution: the standard choice (dashed), 
    ``closest $m_t$'' (solid),
    ``same $m_t$'' (dot-dashed), and the one giving the ``best $m_t$''. 
    Left: particle level, right: \Delphes{} detector level obtained using the ATLAS card.
    Ratios to the standard option are provided in lower panels, the yellow band indicating the statistical uncertainty in the denominator.
    Top (bottom) plots are in the logarithmic (linear) scale.
  }
\label{pst:mt_nupz_study5_match}
\end{figure}

In order to further improve the correlation between the detector and particle levels, current HEP experiments also restrict the analysis phase-space to events where corresponding objects forming the pseudo-top quarks (i.e. the lepton, light jets and $b$-tagged jets) are well angularly matched between the particle and detector levels, using usually a~$\Delta R$ cut of 0.02 for leptons and 0.35 for jets. This leads to much more diagonal migration matrices, as can be seen in Fig.~\ref{pst:migra_study6_ptcl_det_match}. The price for this is an additional matching efficiency of the order of 0.5--0.7 which needs to be compensated for using a~dedicated bin-by-bin correction, while the advantage is that the migration matrix then accounts only for resolution and not for combinatorial effects.
In particular, for the ``best $m_t$'' case, the matching condition between the two $b$-jets had to be relaxed in order to allow for the swap of the $b$-jets, as the strict assignment was otherwise only about 20\% efficient.
The performance of the algorithms on the line shape of the leptonic pseudo-top mass as shown in Fig.~\ref{pst:mt_nupz_study5_match} is similar to the case without the matching requirement (Fig.~\ref{pst:mt_nupz_study5}).
Correlations between the particle and detector levels for the case of matched events are summarized in~Tab.~\ref{tab:corrs:particle_detector_match}, with the highlighted best performing algorithm.

%

%
%

\begin{table}[p]
  \centering
  \begin{tabular}{l|llll} \hline\hline
    observable         & standard & closest $m_t$ & same $m_t$ & best $m_t$ \\ \hline
$\mtt$ & \textbf{0.77} & 0.73 & 0.67 & 0.70 \\
$\ytt$ & \textbf{0.94} & \textbf{0.93} & 0.90 & 0.92 \\
$\deltatt$ & \textbf{0.73} & \textbf{0.72} & 0.67 & 0.62 \\
$\abscosthetastar$ & \textbf{0.68} & 0.66 & 0.64 & 0.52 \\
$\ytlep$ & \textbf{0.97} & 0.94 & 0.86 & 0.88 \\
$\Yboost$ & \textbf{0.86} & 0.84 & 0.80 & 0.80 \\
$\Chittbar$ & \textbf{0.75} & 0.71 & 0.71 & 0.64 \\
    \hline\hline
  \end{tabular}
  \caption{Correlation coefficients of the migration matrices between the particle and \Delphes{} detector levels for different observables (with largest values, within 1\%, highlighted in bold) and various ways to reconstruct the pseudo-$\ttbar$ related observables.}
  \label{tab:corrs:particle_detector}
\end{table}

\begin{table}[p]
  \centering
  \begin{tabular}{l|llll} \hline\hline
    observable         & standard & closest $m_t$ & same $m_t$ & best $m_t$ \\ \hline
$\mtt$ & \textbf{0.96} & 0.94 & \textbf{0.95} & 0.93 \\
$\ytt$ & \textbf{0.99} & \textbf{0.99} & \textbf{0.99} & \textbf{0.98} \\
$\deltatt$ & \textbf{0.98} & 0.96 & 0.96 & 0.87 \\
$\abscosthetastar$ & \textbf{0.95} & 0.92 & \textbf{0.94} & 0.81 \\
$\ytlep$ & \textbf{0.99} & \textbf{0.98} & 0.97 & 0.94 \\
$\Yboost$ & \textbf{0.98} & \textbf{0.97} & \textbf{0.97} & 0.93 \\
$\Chittbar$ & \textbf{0.97} & 0.93 & 0.95 & 0.87 \\
    \hline\hline
  \end{tabular}
  \caption{Correlation coefficients of the migration matrices between the particle and \Delphes{} detector levels for matched events for different observables (with largest values, within 1\%, highlighted in bold) and various ways to reconstruct the pseudo-$\ttbar$ related observables.}
  \label{tab:corrs:particle_detector_match}
\end{table}

In addition, a~comparison to parton-level top quarks was performed, taking the last top quarks in the \Pythia{}8 parton chain, corresponding to top quarks after the final state radiation.
For simplicity, the leptonic top quark at the parton level is taken as the one angularly closer to the particle or detector level leptonic pseudo-top.
Migration matrices between the parton and particle, and parton and detector levels were studied with the following observations.

\begin{figure}[!pt]
  \includegraphics[width=1.00\textwidth]{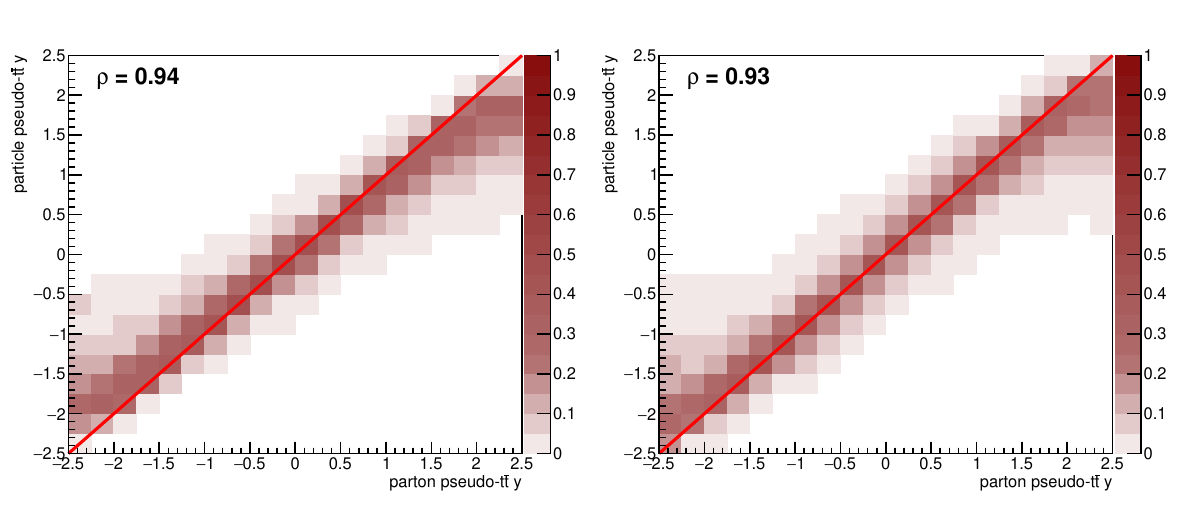}
  \\
  \includegraphics[width=1.00\textwidth]{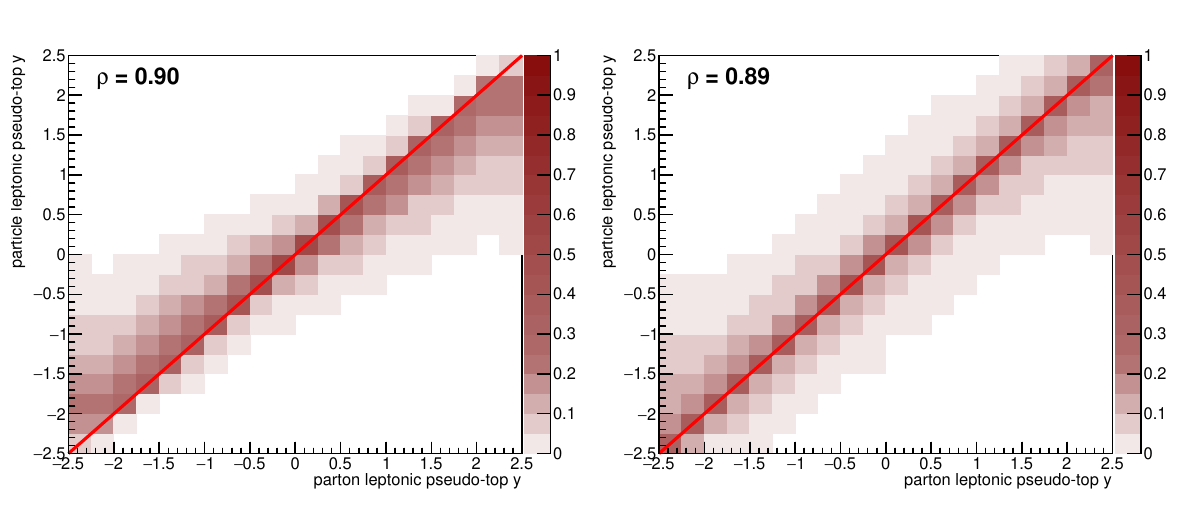}
  \caption{Migration matrices between the parton and particle levels for the 
    pseudo-\ttbar{} rapidity (top)
    and the leptonic pseudo-top rapidity (bottom)
    obtained using the ATLAS card; for different choices of the neutrino $p_z$ solution based on the $m_{\ell\nu}= m_W$ condition: the standard choice (left)
    and the ``closest $m_t$'' (right). 
    The solid bold line is the diagonal, $\rho$ stands for the correlation coefficient evaluated before the normalization of columns.}
\label{pst:migra_study6_parton_ptcl}
\end{figure}

The correlation between the parton and particle levels is shown in Fig.~\ref{pst:migra_study6_parton_ptcl} where only a slight decorrelation is observed for the novel ``closest $m_t$'' method. 
The resulting correlation coefficients for all the studied spectra between the parton and particle or detector levels for more variables are summarized in Tab.~\ref{tab:corrs:parton_particle} or Tab.~\ref{tab:corrs:parton_detector}, respectively. 

\begin{figure}[!pt]
\includegraphics[width=1.00\textwidth]{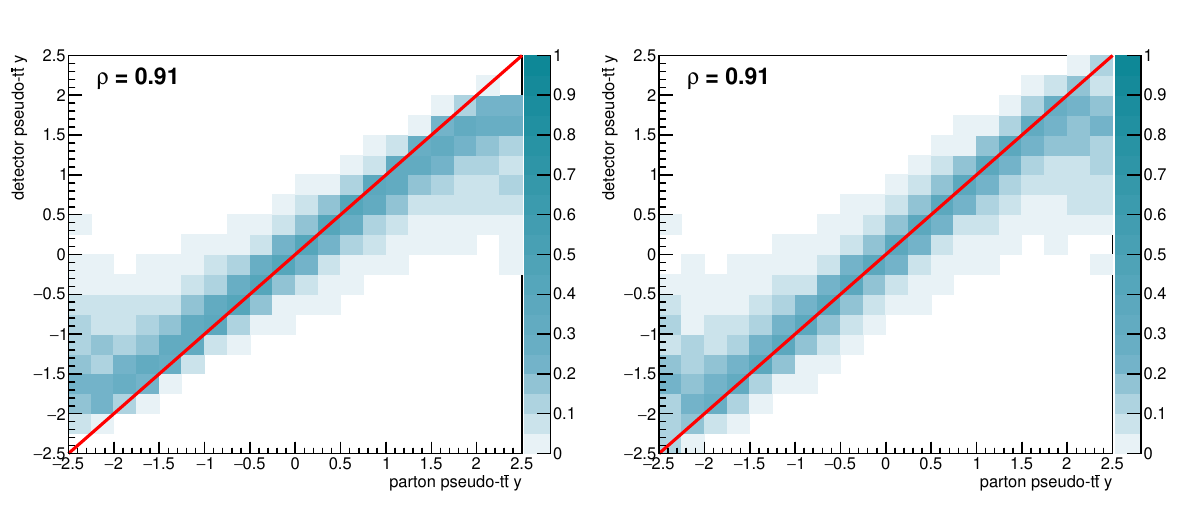}
\\
\includegraphics[width=1.00\textwidth]{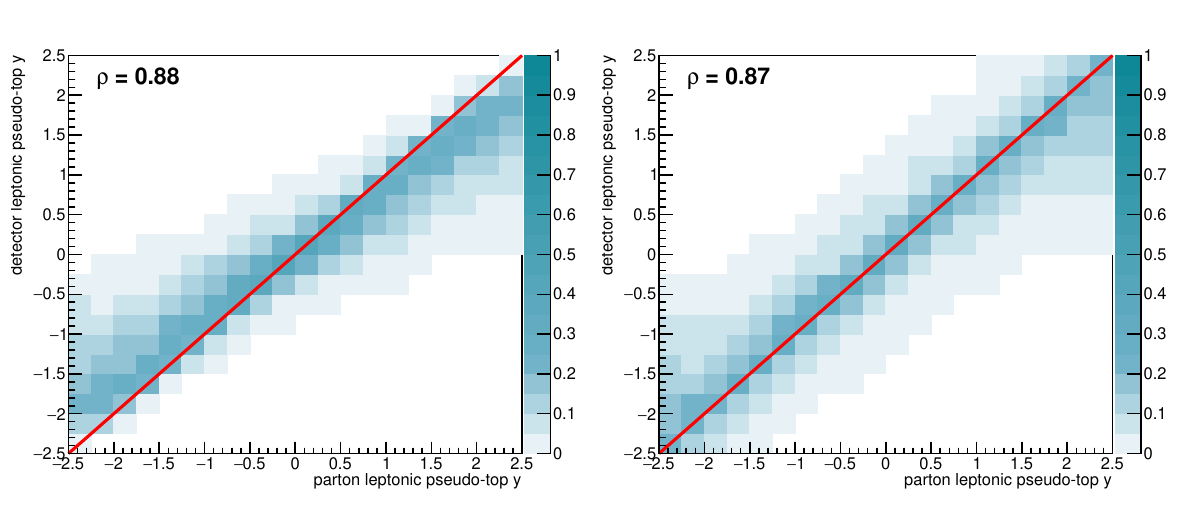}
  \caption{Migration matrices between the parton and the \Delphes{} detector levels for the 
    pseudo-\ttbar{} rapidity (top)
    and the leptonic pseudo-top rapidity (bottom) 
    obtained using the ATLAS card; for different choices of the neutrino $p_z$ solution based on the $m_{\ell\nu}= m_W$ condition: the standard choice (left) 
     and the ``closest $m_t$'' (right). 
    The solid bold line is the diagonal, $\rho$ stands for the correlation coefficient evaluated before the normalization of columns.}
\label{pst:migra_study6_parton_det}
\end{figure}

It can be observed that the correlation between the parton and \Delphes{} detector level is worse for the pseudo-\ttbar{} mass using the ``same $m_t$'' method compared to the standard one, but all correlations are very similar for the standard and the ``closest $m_t$'' methods.
Still, the improved and more careful treatment of the rapidity of the neutrino in the ``closest $m_t$'' method leads to the removal of the ``tilt'' in migration matrices of the rapidities of the pseudo-$\ttbar$ as well as the leptonic pseudo-top compared to the standard method, and 
the parton-to-detector level correspondence is thus more linear (Fig.~\ref{pst:migra_study6_parton_det}).

No tilt observed in migration between the particle and detector levels means the rapidities are similarly biased for these two levels compared to the parton level, as can also be checked in bottom plots of Fig.~\ref{pst:migra_study6_parton_ptcl}.
As the rapidities are used in fits of parton distribution functions (PDF), the ``compression'' of the rapidities of the top quark and the $\ttbar{}$ system using the standard reconstruction method possibly dilutes the information and diminishes the potential to constrain the PDF functions, while it could be partially recovered using the proposed ``closest $m_t$'' method.

\begin{table}[p]
\centering
  \begin{tabular}{l|llll} \hline\hline
    observable        & standard & closest $m_t$ & same $m_t$ & best $m_t$ \\ \hline
$\mtt$ & \textbf{0.75} & \textbf{0.74} & 0.64 & 0.71 \\
$\ytt$ & \textbf{0.94} & \textbf{0.93} & 0.90 & 0.92 \\
$\deltatt$ & \textbf{0.68} & \textbf{0.68} & 0.58 & 0.59 \\
$\abscosthetastar$ & 0.48 & \textbf{0.50} & 0.36 & 0.46 \\
$\ytlep$ & \textbf{0.90} & \textbf{0.89} & 0.78 & 0.87 \\
$\Yboost$ & \textbf{0.83} & 0.80 & 0.78 & 0.79 \\
$\Chittbar$ & \textbf{0.58} & \textbf{0.57} & 0.47 & 0.51 \\
    \hline\hline
  \end{tabular}
  \caption{Correlation coefficients for the migration matrices between the parton and particle levels for different observables  (with largest values, within 1\%, highlighted in bold) and various ways to reconstruct the pseudo-$\ttbar$ related observables.}
  \label{tab:corrs:parton_particle}
\end{table}

\begin{table}[p]
  \centering
  \begin{tabular}{l|llll} \hline\hline
    observable         & standard & closest $m_t$ & same $m_t$ & best $m_t$ \\ \hline
$\mtt$ & \textbf{0.65} & 0.63 & 0.53 & 0.60 \\
$\ytt$ & \textbf{0.91} & \textbf{0.91} & 0.85 & \textbf{0.90} \\
$\deltatt$ & \textbf{0.69} & \textbf{0.68} & 0.59 & 0.62 \\
$\abscosthetastar$ & \textbf{0.46} & \textbf{0.46} & 0.42 & \textbf{0.46} \\
$\ytlep$ & \textbf{0.89} & \textbf{0.88} & 0.75 & 0.87 \\
$\Yboost$ & \textbf{0.79} & \textbf{0.78} & 0.71 & 0.76 \\
$\Chittbar$ & \textbf{0.59} & 0.56 & 0.53 & 0.54 \\
\hline\hline
  \end{tabular}
  \caption{Correlation coefficients for the migration matrices between the parton and \Delphes{} detector levels for different observables (with largest values, within 1\%, highlighted in bold) and various ways to reconstruct the pseudo-$\ttbar$ related observables.}
  \label{tab:corrs:parton_detector}
\end{table}

%
%

%
%

\subsection{Unfolding performance}
\label{sec:unf}

\begin{figure}[!pt]
  \includegraphics[width=0.45\textwidth]{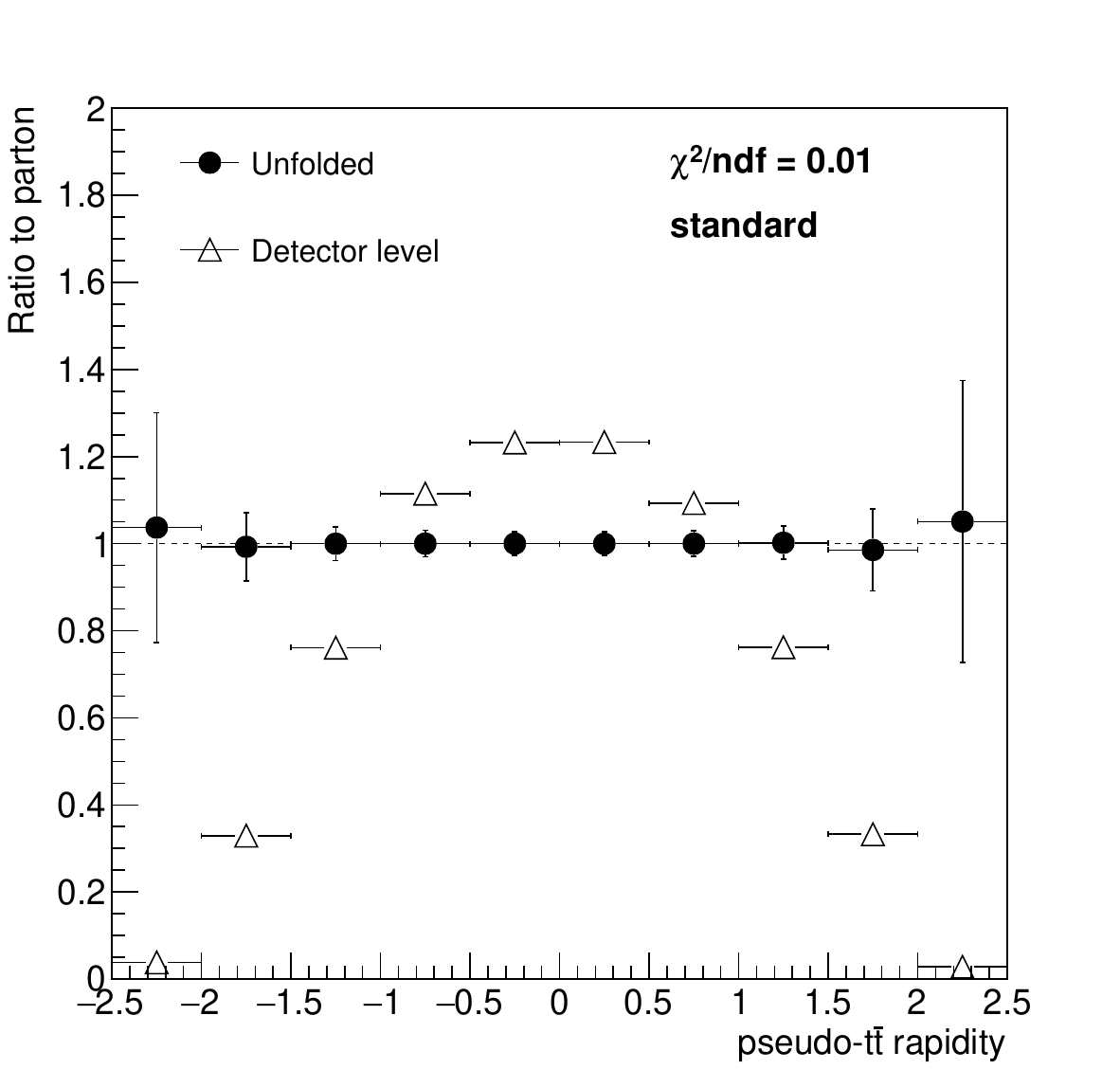} 
  \includegraphics[width=0.45\textwidth]{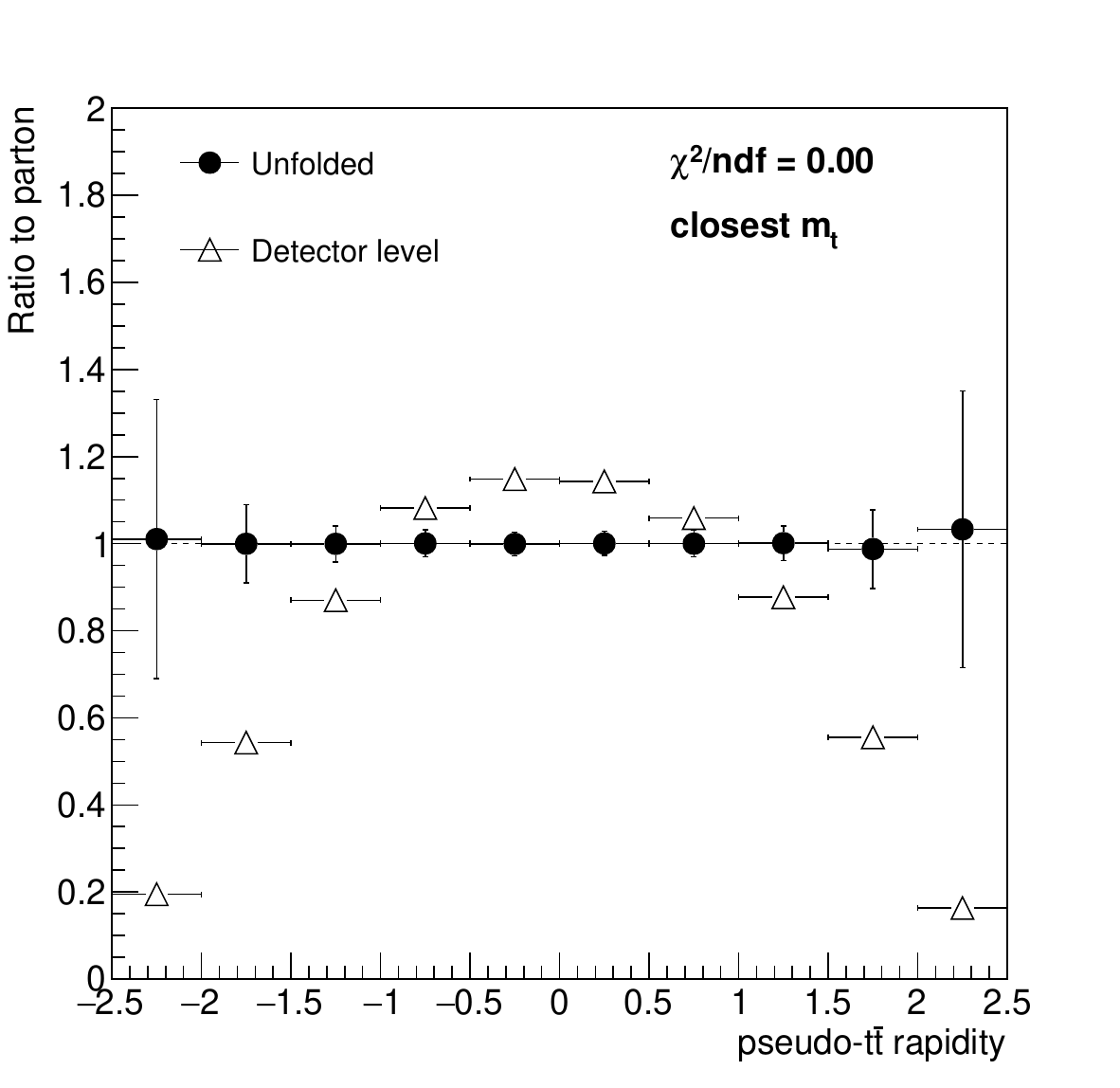} 
\\
  \caption{Unfolding closure test (ratio of the unfolded \Delphes{} detector level to the parton level) for the
    pseudo-\ttbar{} rapidity 
    for different choices of the neutrino $p_z$ solution based on the $m_{\ell\nu}= m_W$ condition: the standard choice (left)
    and the ``closest $m_t$'' (right).
  }
\label{pst:unf_study_parton_closure}
\end{figure}

In order to check the performance of correcting the detector-level spectra for resolution effects (unfolding), a Python implementation~\cite{fbu_py} of the Fully Bayesian Unfolding technique~\cite{FBU} was used to unfold the rapidity spectra of the pseudo-$\ttbar$ system to the parton level. In detail, the \Delphes{} detector level spectrum from the projection of the response matrix was used as input pseudo-data and
comparison was made after unfolding to the original parton-level spectrum from the projection of the response matrix on the other axis.
It was checked that the unfolded posterior distributions are very well Gaussian and the posterior mean was taken as the unfolded result in each bin.
Results in Fig.~\ref{pst:unf_study_parton_closure} show, besides the largely more central detector-level spectrum for the standard neutrino $p_z$ choice (empty triangles), that a perfect closure (full points) is reached for both standard and ``closest $m_t$'' choice in terms of the $\chi^2/\mathrm{ndf} \leq 0.01$, by comparing the unfolded histogram divided by the parton-level spectrum to unity. 
Thus the two options are equivalent in unfolding performance in terms of a closure test within the same sample.

In reality, however, more stringent unfolding tests are needed as the spectrum in data is not the same as in simulation.
Different simulation samples lead to different migration matrices and efficiency corrections, which are thus model-dependent.
Larger difference between spectra at the detector and parton level can lead to unfolding non-closure which needs to be treated as a systematics.

\begin{figure}[!pt]
  \includegraphics[width=0.95\textwidth]{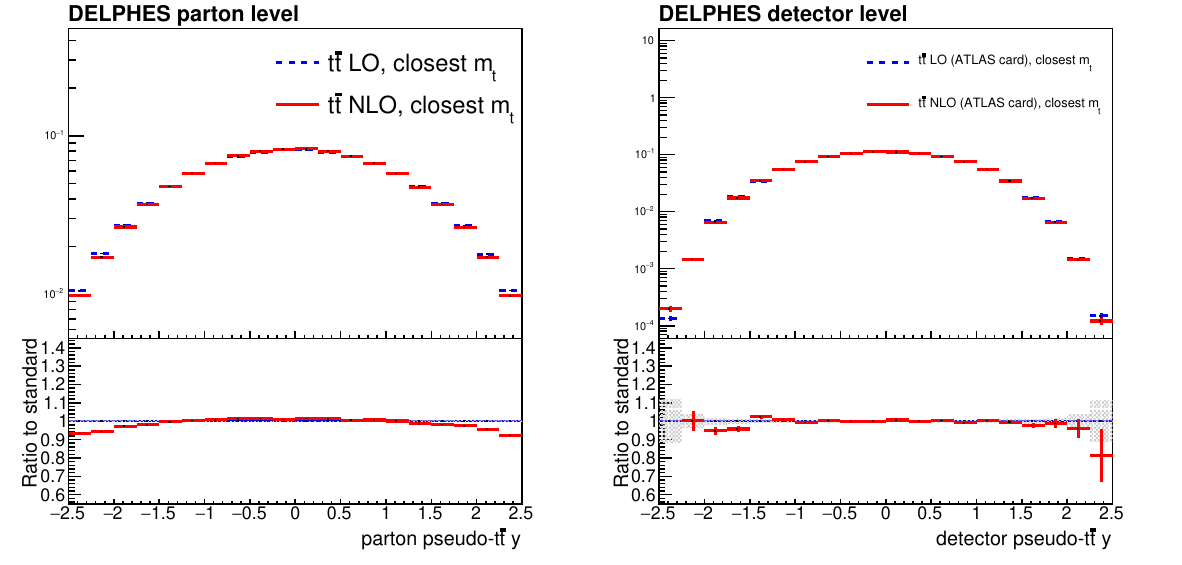} 
  \caption{Comparison of the LO (dashed) and NLO (solid) shapes of the pseudo-\ttbar{} rapidity distribution at the parton (left) and \Delphes{} detector (right) level obtained using the ATLAS card and for the ``closest $m_t$'' option at the detector level.}
\label{pst:study_parton_det_LO_NLO_diff}
\end{figure}

\begin{figure}[!pt]
  \includegraphics[width=0.45\textwidth]{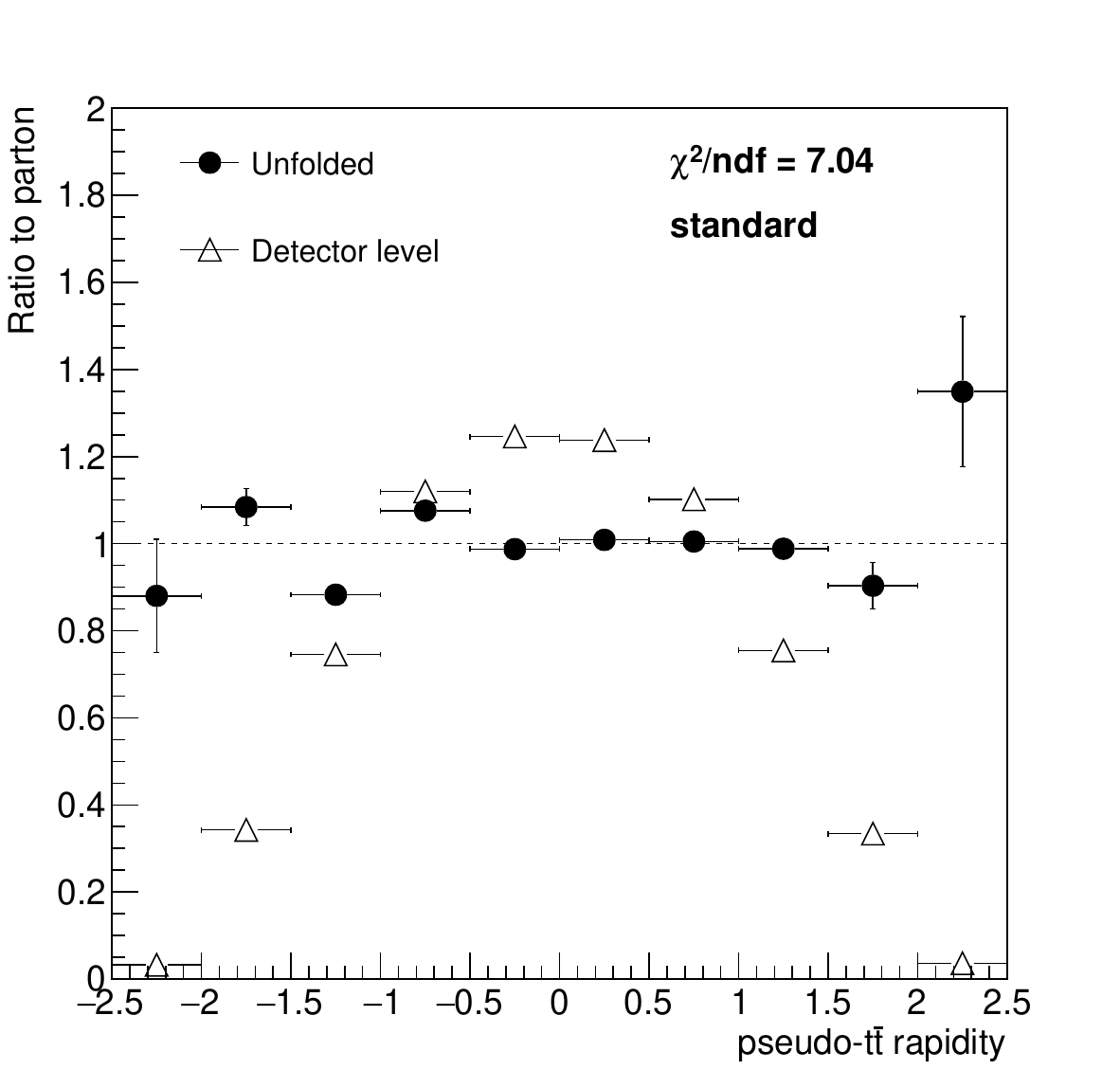} 
  \includegraphics[width=0.45\textwidth]{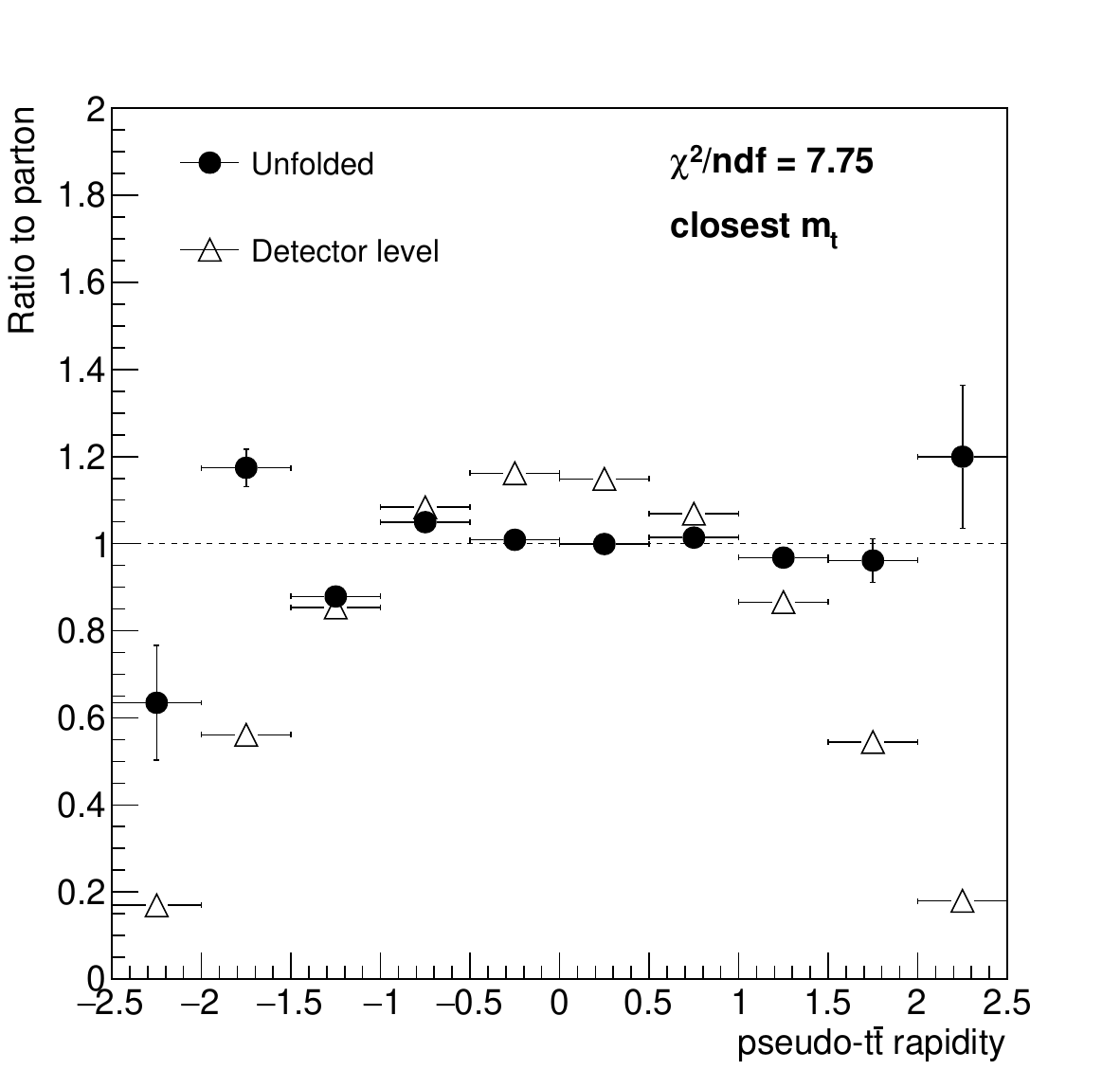} 
  \caption{Unfolding the LO $\ttbar$ sample using the migration matrix from the NLO $\ttbar$ sample. Closure test (ratio of the unfolded \Delphes{} detector level to the parton level) for the
    pseudo-\ttbar{} rapidity 
    for different choices of the neutrino $p_z$ solution based on the $m_{\ell\nu}= m_W$ condition: the standard choice (left)
    and the ``closest $m_t$'' (right).
  }
\label{pst:unf_study_parton_det_gensyst}
\end{figure}

\begin{figure}[!pt]
  \includegraphics[width=0.45\textwidth]{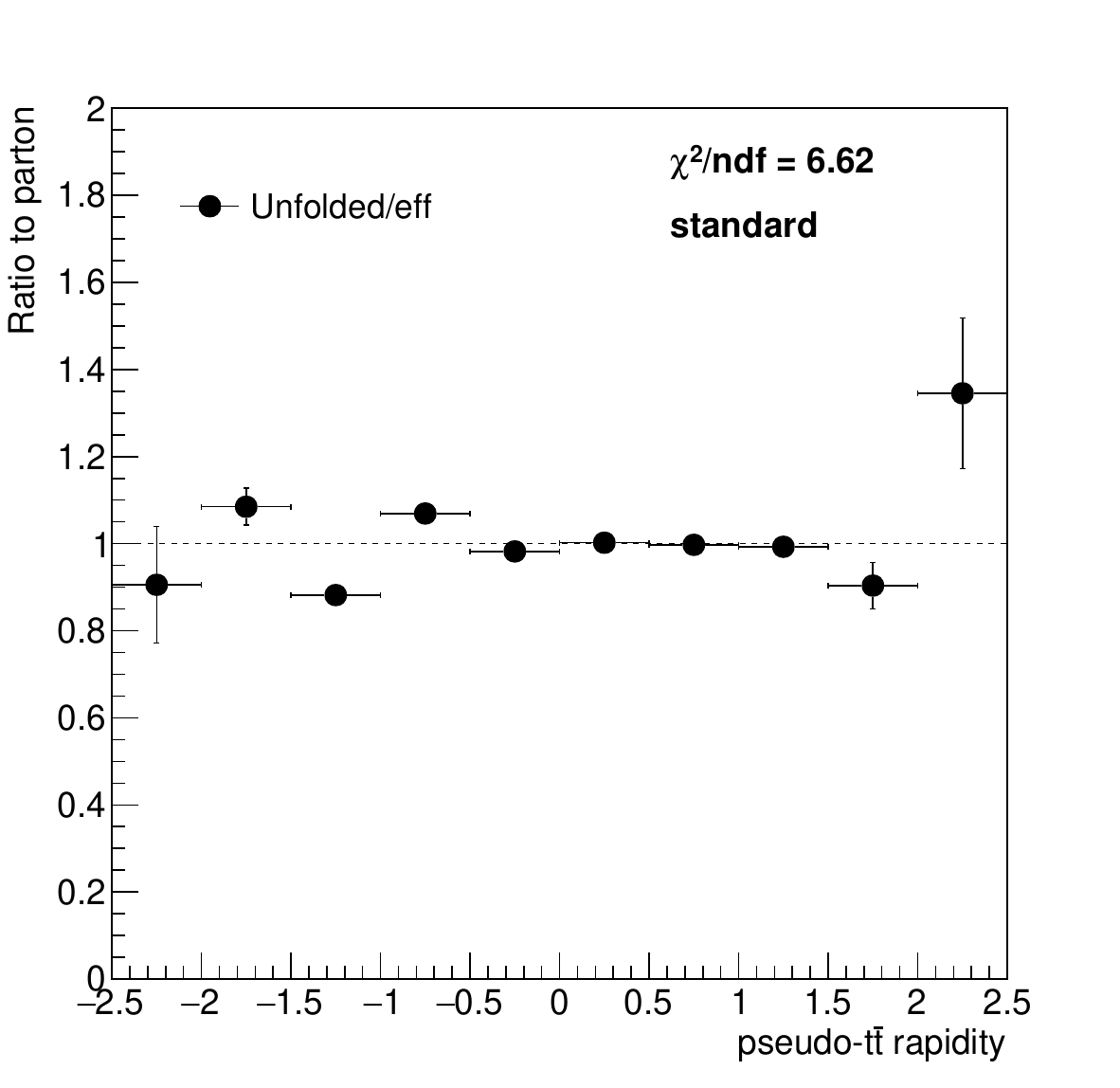} 
  \includegraphics[width=0.45\textwidth]{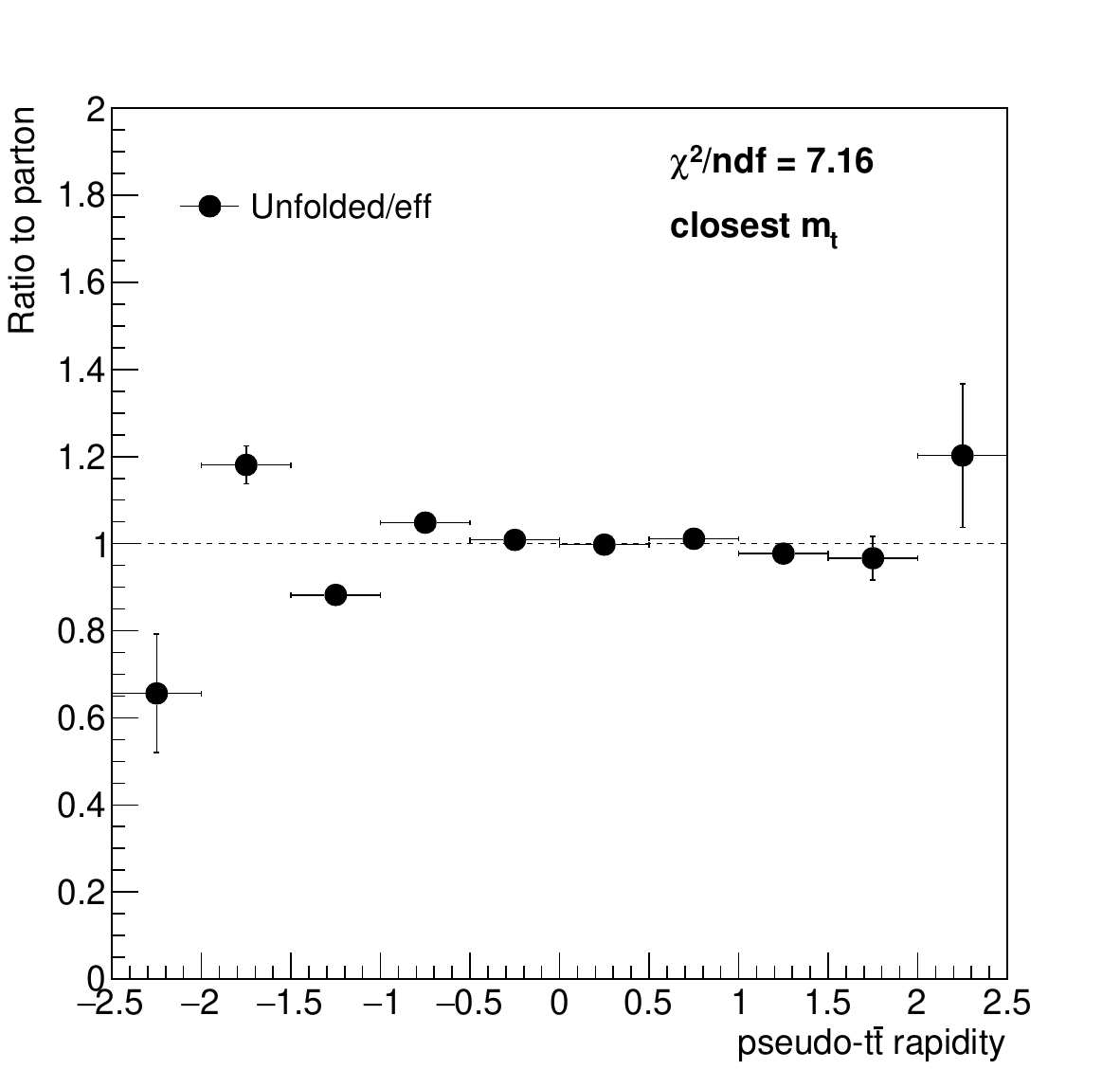} 
  \caption{Unfolding the LO $\ttbar$ sample to the full partonic phase-space using the migration matrix and the efficiency correction from the NLO $\ttbar$ sample. Normalized closure test (ratio of the normalized unfolded \Delphes{} detector level to the normalized parton level) for the
    pseudo-\ttbar{} rapidity 
    for different choices of the neutrino $p_z$ solution based on the $m_{\ell\nu}= m_W$ condition: the standard choice (left)
    and the ``closest $m_t$'' (right).
  }
\label{pst:unf_study_parton_det_gensyst_eff}
\end{figure}

The following tests are motivated by one of the dominant systematics uncertainties in real measurements which is often due to the choice of the $\ttbar$ generator to derive the corrections.
A more realistic closure test was thus performed using the LO $\ttbar$ sample and unfolding it using the migration matrix derived from the NLO $\ttbar$ sample. The difference between the spectra at the LO and NLO is depicted in~Fig.~\ref{pst:study_parton_det_LO_NLO_diff}.
The unfolding closure test without scaling to the full partonic phase space is shown in Fig.~\ref{pst:unf_study_parton_det_gensyst} while the full closure test, i.e. including the efficiency correction to the full partonic phase-space, is shown in~Fig.~\ref{pst:unf_study_parton_det_gensyst_eff}. Due to the fact that the efficiency derived using the NLO sample is about 3\% higher than that of the LO sample because of kinematics, the closure test was performed between normalized distributions and the number of degrees of freedom (ndf) was lowered by one.
In both cases, a~comparable performance in terms of the $\chi^2$ test can be observed.

\subsection{Spectra comparison}
\begin{figure}[p]
      \includegraphics[width=1.00\textwidth]{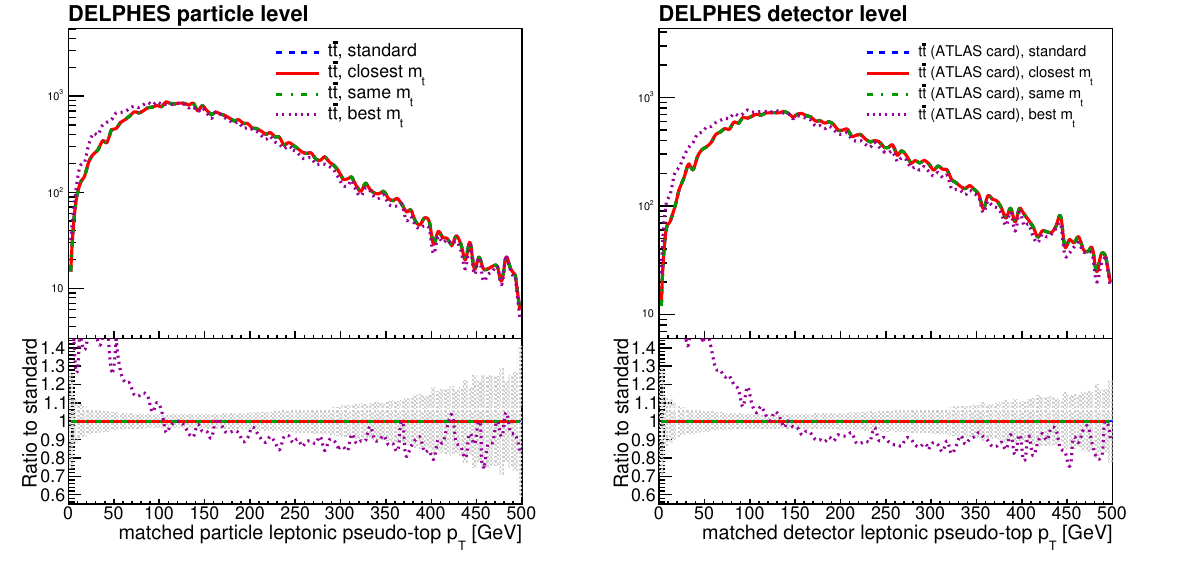} 
\\
      \includegraphics[width=1.00\textwidth]{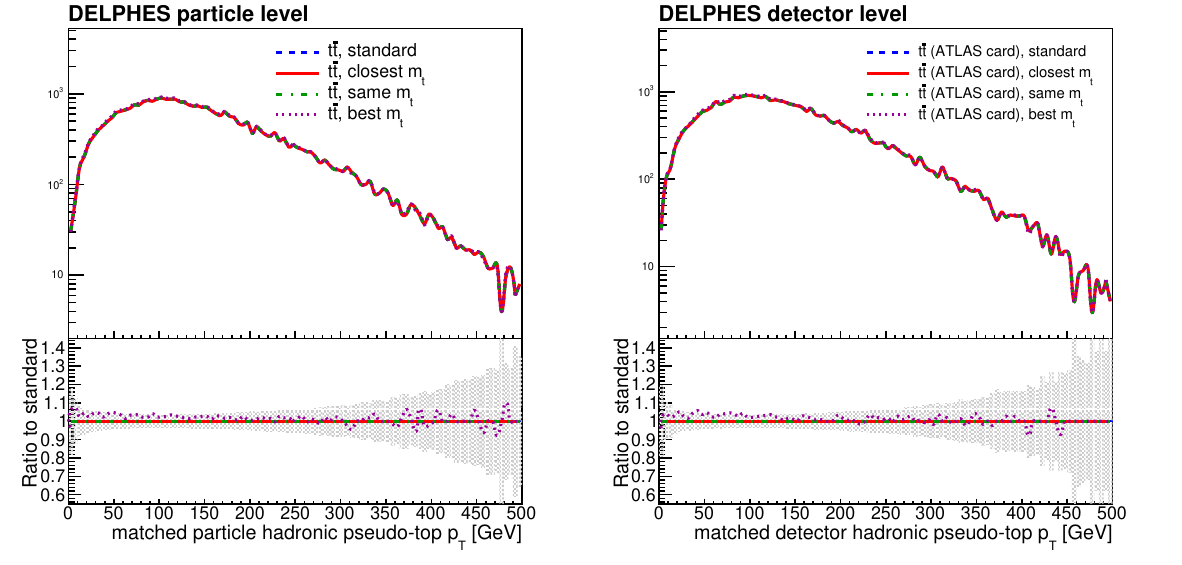} 
 \caption{Distributions of the leptonic (top) and hadronic (bottom) pseudo-top quark transverse momentum for matched events 
   for different choices of the neutrino $p_z$ solution: the standard choice (dashed),
   ``closest $m_t$'' (solid), 
   ``same $m_t$'' (dot-dashed), and the 
   ``best $m_t$'' (dotted). 
    Left: particle level, right: \Delphes{} detector level obtained using the ATLAS card.
    Ratios to the standard option are provided in lower panels, the yellow band indicating the statistical uncertainty in the denominator.
  }
\label{pst:mt_nupz_study6_match}
\end{figure}

\begin{figure}[p]
  \includegraphics[width=1.00\textwidth]{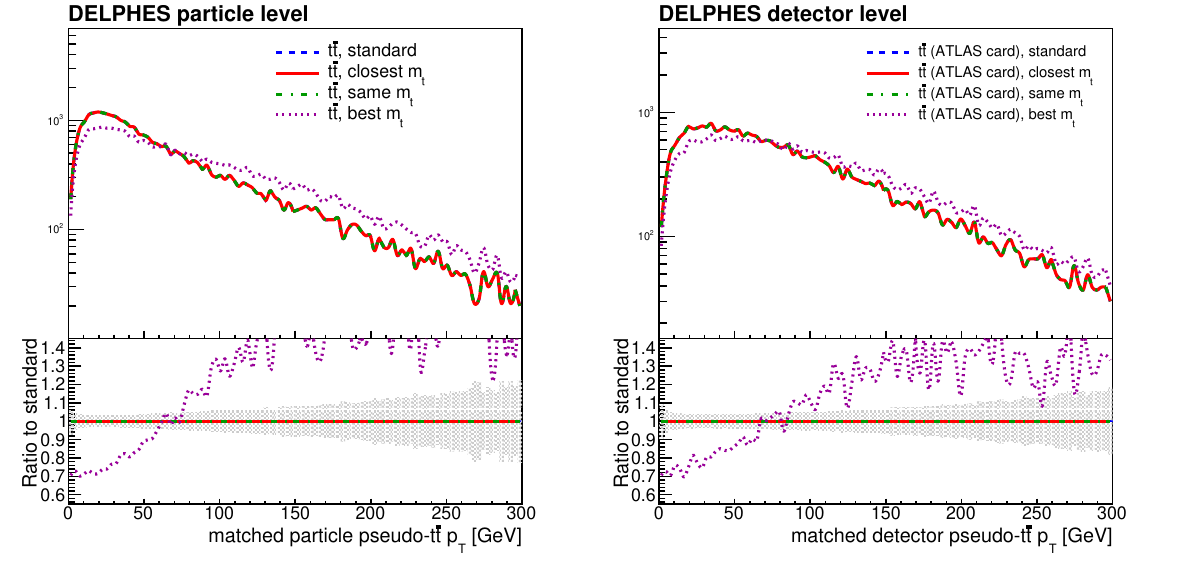} 
\\
  \includegraphics[width=1.00\textwidth]{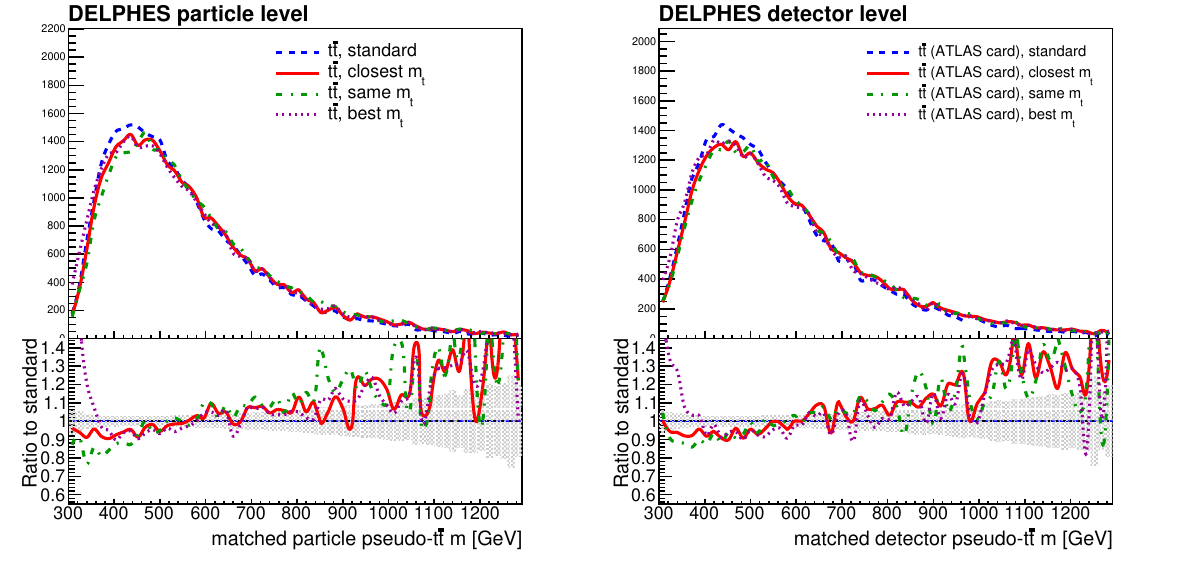} 

  \caption{Distributions the pseudo-$\ttbar$ transverse momentum (top) and mass (bottom) for matched events 
    for different choices of the neutrino $p_z$ solution: the standard choice (dashed),
   ``closest $m_t$'' (solid), 
   ``same $m_t$'' (dot-dashed), and the 
     ``best $m_t$'' (dotted). 
    Left: particle level, right: \Delphes{} detector level obtained using the ATLAS card.
    Ratios to the standard option are provided in lower panels, the yellow band indicating the statistical uncertainty in the denominator.
  }
\label{pst:mt_nupz_study7_match}
\end{figure}

\begin{figure}[p]
  \includegraphics[width=1.00\textwidth]{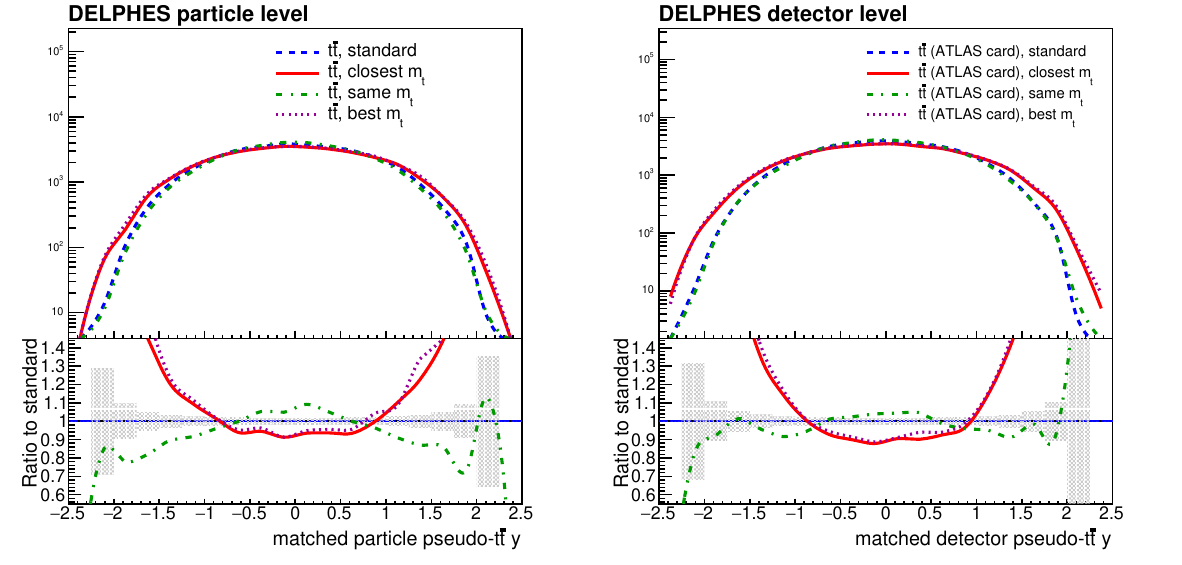} 
\\ 
  \includegraphics[width=1.00\textwidth]{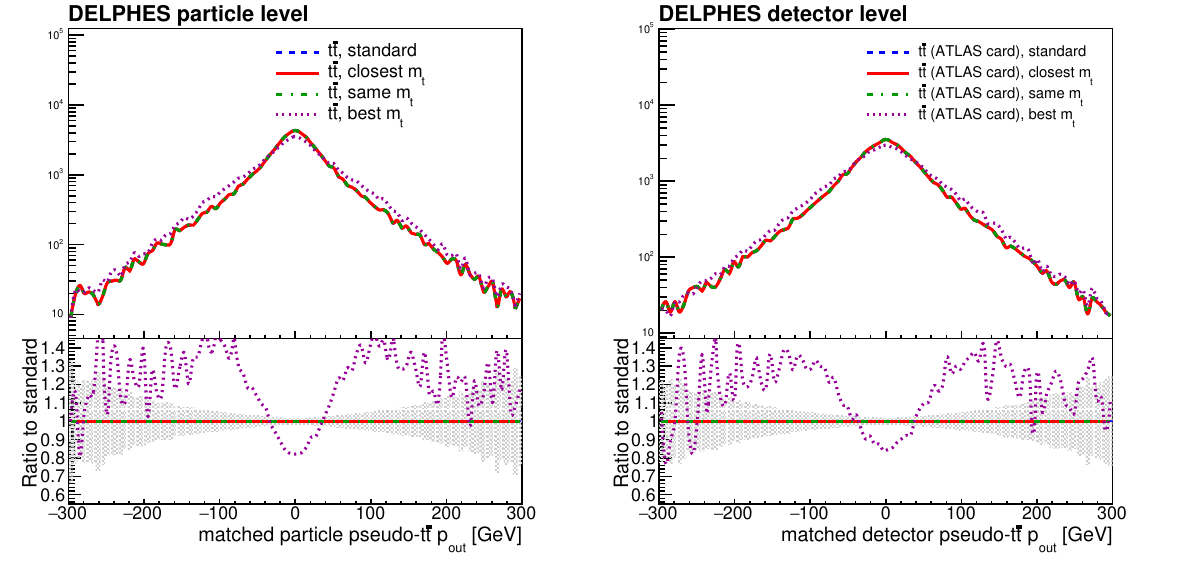} 
  \caption{Distributions of the pseudo-$\ttbar$ rapidity (top) and the $\Pout$ (bottom) variable for matched events
        for different choices of the neutrino $p_z$ solution: the standard choice (dashed),
   ``closest $m_t$'' (solid), 
   ``same $m_t$'' (dot-dashed), and the 
     ``best $m_t$'' (dotted). 
    Left: particle level, right: \Delphes{} detector level obtained using the ATLAS card.
    Ratios to the standard option are provided in lower panels, the yellow band indicating the statistical uncertainty in the denominator.
  }
\label{pst:mt_nupz_study8_match}
\end{figure}

\begin{figure}[p]
  \includegraphics[width=1.00\textwidth]{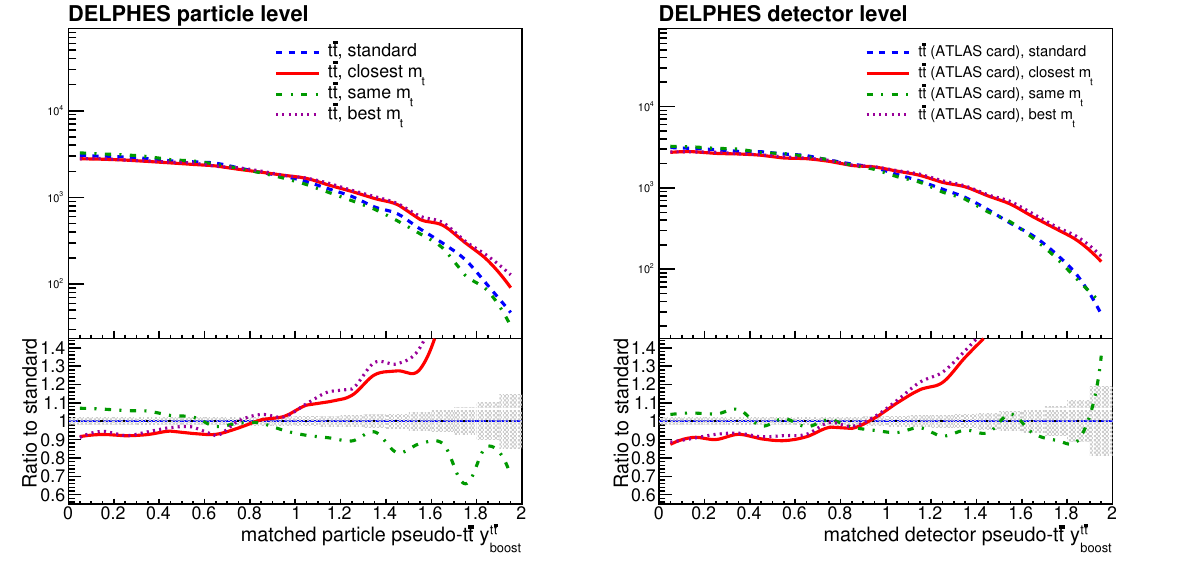} 
\\
  \includegraphics[width=1.00\textwidth]{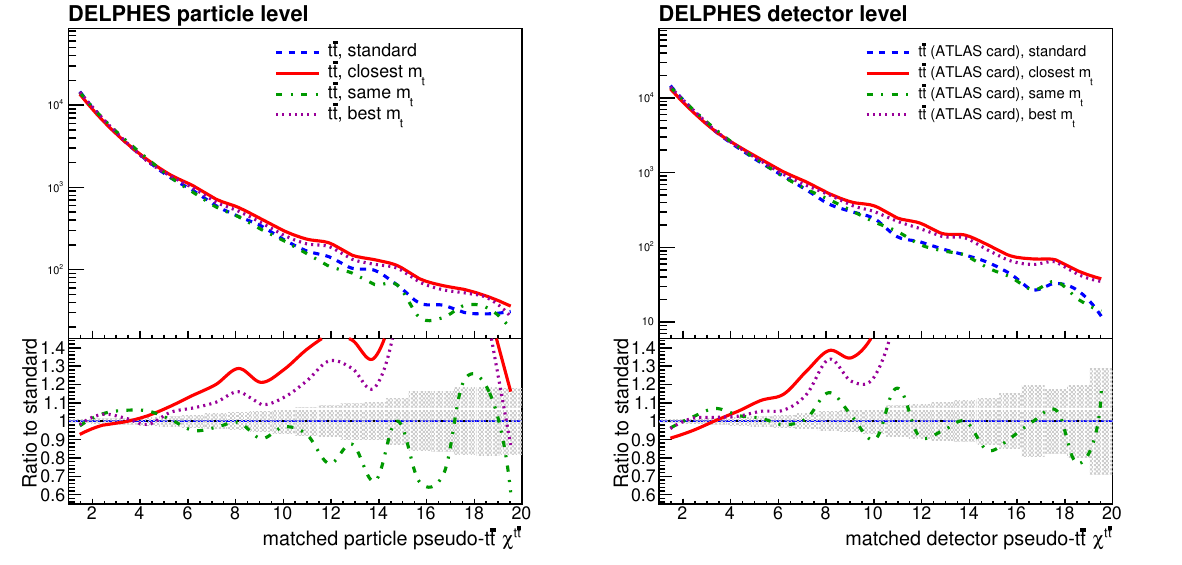} 
  \caption{Distributions of the $y_{\mathrm{boost}}^{\ttbar}$ (top) and $\chi^{\ttbar}$ (bottom) variables 
 for different choices of the neutrino $p_z$ solution: the standard choice (dashed),
   ``closest $m_t$'' (solid), 
   ``same $m_t$'' (dot-dashed), and the 
     ``best $m_t$'' (dotted). 
    Left: particle level, right: \Delphes{} detector level obtained using the ATLAS card.
    Ratios to the standard option are provided in lower panels, the yellow band indicating the statistical uncertainty in the denominator.
  }
\label{pst:mt_nupz_study9_match}
\end{figure}

Additional information is provided by the comparison of shapes of several physics observables used in applications like tuning; these are shown in~Figs.~\ref{pst:mt_nupz_study6_match}--\ref{pst:mt_nupz_study9_match} which show the spectra at particle and detector levels with the angular matching required between objects forming the pseudo-tops at the two levels (see Sec.~\ref{sec:match}). For spectra of transverse momenta of leptonic and hadronic pseudo-tops (Fig.~\ref{pst:mt_nupz_study6_match}) and the $\ttbar$ system (Fig.~\ref{pst:mt_nupz_study7_match}), and of the out-of-plane momentum $p_\mathrm{out}$ (Fig.~\ref{pst:mt_nupz_study8_match}) all solutions are equivalent except for the ``best $m_t$'' case where large slope changes are observed, disfavouring this option, however well-motivated it had seemed in allowing also the $b$-jets swap (thus affecting also the hadronic-top and $\pt$-related quantities). The standard and ``same $m_t$'' choices lead to unnaturally more central rapidities of the leptonic pseudo-top and of the $\ttbar{}$ system (Figs.~\ref{pst:mt_nupz_study7_match}--\ref{pst:mt_nupz_study8_match}). Interestingly, large slope differences are also observed for higher values of the $\Yboost{}$ and $\Chittbar{}$ variables (Fig.~\ref{pst:mt_nupz_study9_match}) which are of interest for new physics searches using top quarks, and a proper choice of the pseudotop algorithm could be done based on the performance of these variables for particular models. But this task is beyond the scope of this study.

\section{Conclusions}

A detailed study of the past, current as well as further modified pseudo-top algorithms and their details used in recent HEP measurements was presented at both the particle and detector levels using $\ttbar{}$ events generated by \MadGraph{} and detector response simulated by \Delphes{}, with particle level analyzed also within the standard \Rivet{} framework. Correlations and unfolding to the parton level were also studied.

Differences are highlighted between the different pseudo-top algorithms in their behaviour especially for the rapidity of objects based on the choice of the neutrino longitudinal momentum from the generally two solutions of the quadratic equation based on the  $m_{\ell\nu}= m_W$ or $m_{t,\mathrm{had}} = m_{t,\mathrm{lep}}$ condition.

An improvement in the pseudo-top algorithm is possible for rapidities of the leptonic pseudo-top and the pseudo-$\ttbar$ system and also seen in the peak the reconstructed leptonic pseudo-top mass when the neutrino $p_z$ choice is done upon the smallest difference of the reconstructed pseudo-top quark masses (the ``closest $m_t$'' case).

Improvement is also checked in terms of the invariant mass of the pseudo-top quark pair for a hypothetical $Z'$ particle of mass of $700\,$GeV and decaying to a $\ttbar$ pair, where a~sharper line is observed at the particle level, indicating better resolution reached in this variable, important for searches for new physics, although the performance at the detector level is smeared due to detector resolution effects.

A summary of pro's and con's of the presented methods is presented in~Table~\ref{tab:pros_cons}.
In particular, the suggested novel ``closest $m_t$'' approach keeps almost the same correlations between detector and particle or parton levels as the standard choice of the neutrino $p_z$, while it has been shown that it provides more realistic spectra (especially less centrally biased rapidities) and outperforms the standard choice in a realistic unfolding test to the parton level, including the efficiency correction. While the ``same $m_t$'' or ``best $m_t$'' methods were motivated in further constraining the leptonic pseudo-top mass (and actually performing better around the peak for the leptonic pseudo-top mass distribution) or allowing the swap of $b$-jets, respectively, they result in undesired tails in the leptonic pseudo-top mass distribution and large slopes in spectra of physics interest. 

In conclusion, the current pseudo-top algorithm used at LHC seems to be sufficient and robust enough for current observables.
Still, improvements in terms of correlations between parton, particle and detector levels could be reached
using the ``closest $m_t$'' method, namely by performing more linearly for the rapidity of the leptonic pseudo-top and the pseudo-\ttbar{} system, though showing a slightly worse, yet comparable, unfolding closure. 
These variables in particular are useful and used in PDF fitting efforts~\cite{Czakon:2016olj}.

Last, \ref{app1} details explicit forms of solutions to quadratic equations for the $p_z^\nu$ problems.

\begin{table}[!h]
  \begin{tabular}{l|l|l|l|l} \hline \hline
                                & standard                    & closest $m_t$                   & same $m_t$                     & best $m_t$ \\ \hline
    Pro's                       & already used,               & better linearity                & sharper $m_t^{\mathrm{lep}}$      & smaller low $m_t^{\mathrm{lep}}$ \\
                                & good general                & for rapidity spectra,           &  peak                          & tail \\ 
                                & performance                 & less central $y_t^{\mathrm{lep}}$,  &                               & \\ 
    &                             & smaller low $m_t^{\mathrm{lep}}$ tail & & \\ 
    &                             & & & \\ \hline

    Con's                       &      not optimized          & slight decorrelation          & too hard                        & modified spectra \\
                                &      for new energies       & for some variables            & high $m_t^{\mathrm{lep}}$          & higher $m_t^{\mathrm{lep}}$   \\ 
                                &                             &                               & tail                               & tail                          \\ \hline \hline
  \end{tabular}
  \caption{A summary table of pro's and con's of the studied pseudo-top algorithms.}
  \label{tab:pros_cons}
\end{table}

\section{Acknowledgements}
The author gratefully acknowledges the support by the project LO1305 of the Ministry of Education, Youth and Sports of the Czech Republic.

\appendix
\section{Analytic solutions to the neutrino $p_z$}
\label{app1}

\subsection{Solution to the $m_{\ell\nu}= m_W$ condition}
The condition $m_{\ell\nu}= m_W$ leads to a quadratic equation for the longitudinal neutrino momentum $p_z^\nu$ with coefficients in standard notation given by
  $$ a = ( E^\ell )^2 - ( p_z^{\ell} )^2$$
  $$ b = -2 \, k^2 \, p_z^{\ell}$$
  $$ c = ( E^\ell \, \slashed{E}_T )^2  - k^4$$
where
  $$ k^2 = \frac12 \left[ ( m_W^2 - ( m^{\ell} )^2 ) \right] + (p_x^{\ell}\, \slashed{E}_x  + p_y^{\ell}\, \slashed{E}_y) $$
and the nature of the solution (complex, one real or two real) is governed by the sign of the usual discriminant $D \equiv b^2 - 4ac$ (here of dimension GeV${}^{6}$).

\subsection{Solution to the $m_{t,\mathrm{had}} = m_{t,\mathrm{lep}}$ condition}
The condition $m_{t,\mathrm{had}} = m_{t,\mathrm{lep}}$ leads to a quadratic equation for the longitudinal neutrino momentum $p_z^\nu$ with coefficients in standard notation given by
    $$ a = 4 \, \left[ (p_z^{\mathrm{sum}})^2 - \Sigma_E^2 \right] $$
    $$ b = 4 \, \Delta m^2 \, p_z^{\mathrm{sum}} $$
    $$ c = \Delta m^2 - 4 \, \slashed{E}_T^2 \, \Sigma_E^2 $$
where
  $$ \Sigma_E^2 \equiv (E^\ell + E^{b_\ell})^2 $$
  $$ \Sigma^2 \equiv (m^{b_\ell})^2 + (m^\ell)^2 - 2 \, \Delta p^2 + 2 \, E^\ell \, E^{b_\ell} - 2 \, \left[ (p^\ell_x + p^{b_\ell}_x ) \, \slashed{E}_x + ( p^\ell_y + p^{b_\ell}_y ) \, \slashed{E}_y \right] $$
  $$ \Delta p^2 = \vec{p}_\ell \cdot \vec{p}_{b_\ell}$$
  $$ p_z^{\mathrm{sum}} \equiv p_z^\ell + p_z^{b_\ell} $$
  $$ \Delta m^2 \equiv m^2_{t,\mathrm{had}} - \Sigma^2 $$
  $$ \slashed{E}_T^2 \equiv (\slashed{E}_x)^2 + (\slashed{E}_y)^2 \,.$$

\subsection{Migration matrices for the discriminants}
\begin{figure}[p]
  \includegraphics[width=0.45\textwidth]{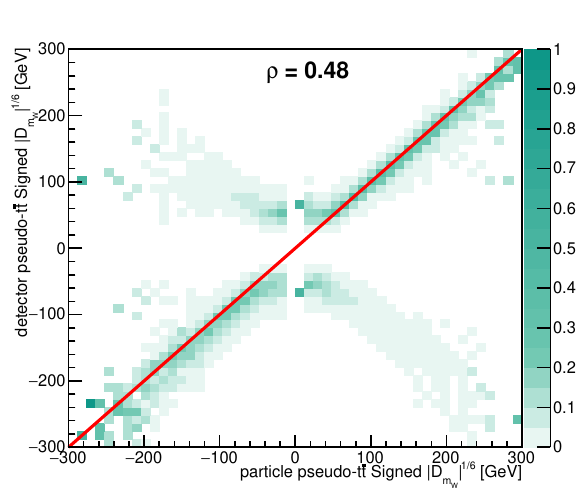} 
  \includegraphics[width=0.45\textwidth]{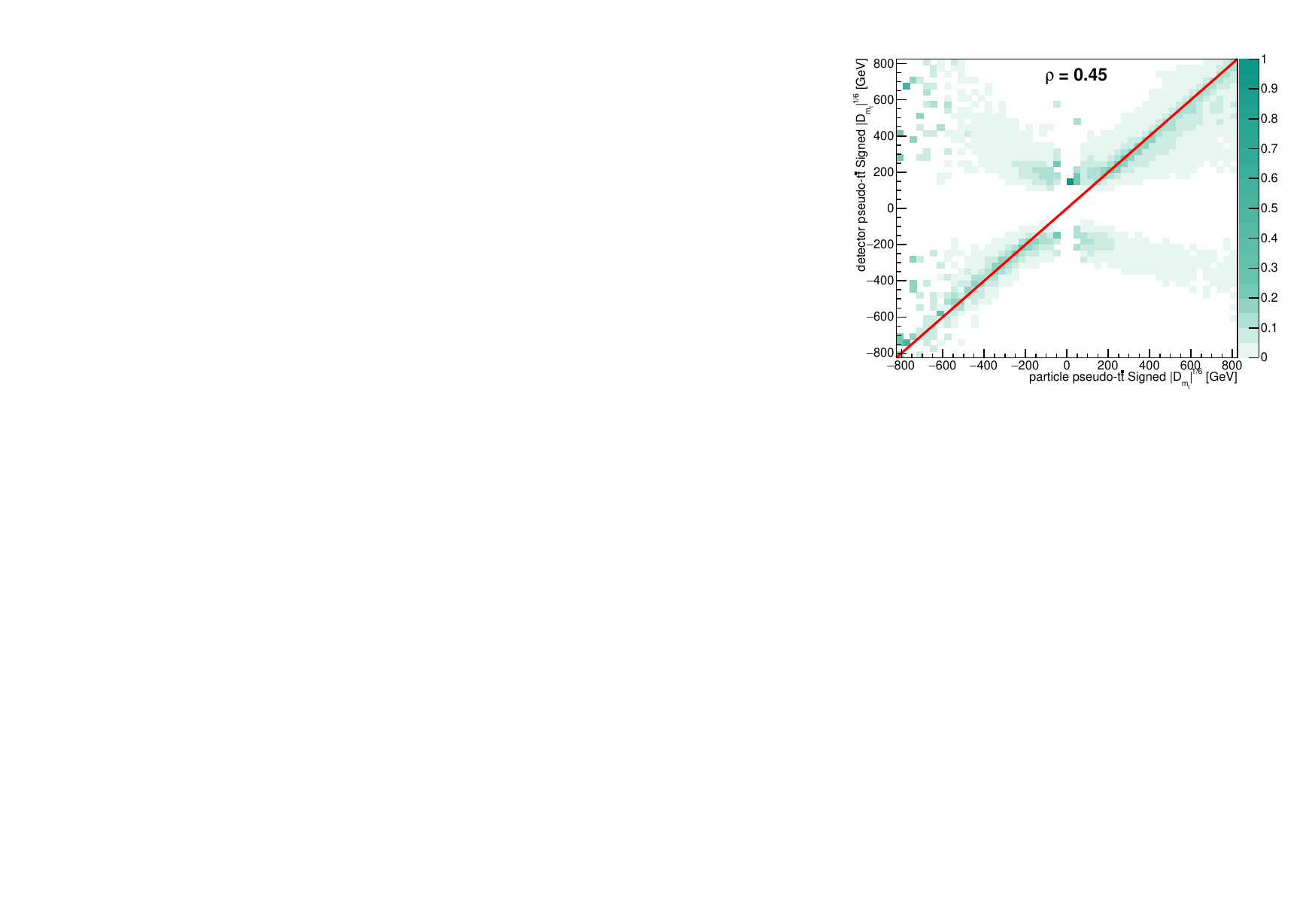} 

\caption{The migration matrix between the particle and the \Delphes{} detector levels for the signed discriminant (to the power of $1/6$ to keep the GeV unit) for the neutrino $p_z$ solution based on the $m_{\ell\nu}= m_W$ condition (left) and based on the $m_{t,\mathrm{had}} = m_{t,\mathrm{lep}}$ condition (right). }
\label{pst:pseudotop_Disc_study1}
\end{figure}

\begin{figure}[p]
    \includegraphics[width=0.45\textwidth]{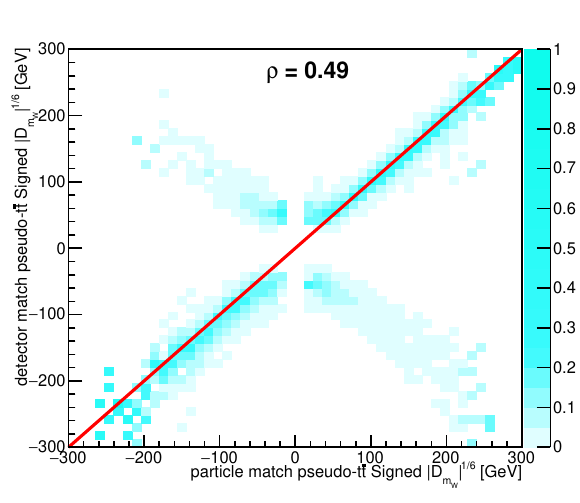} 
    \includegraphics[width=0.45\textwidth]{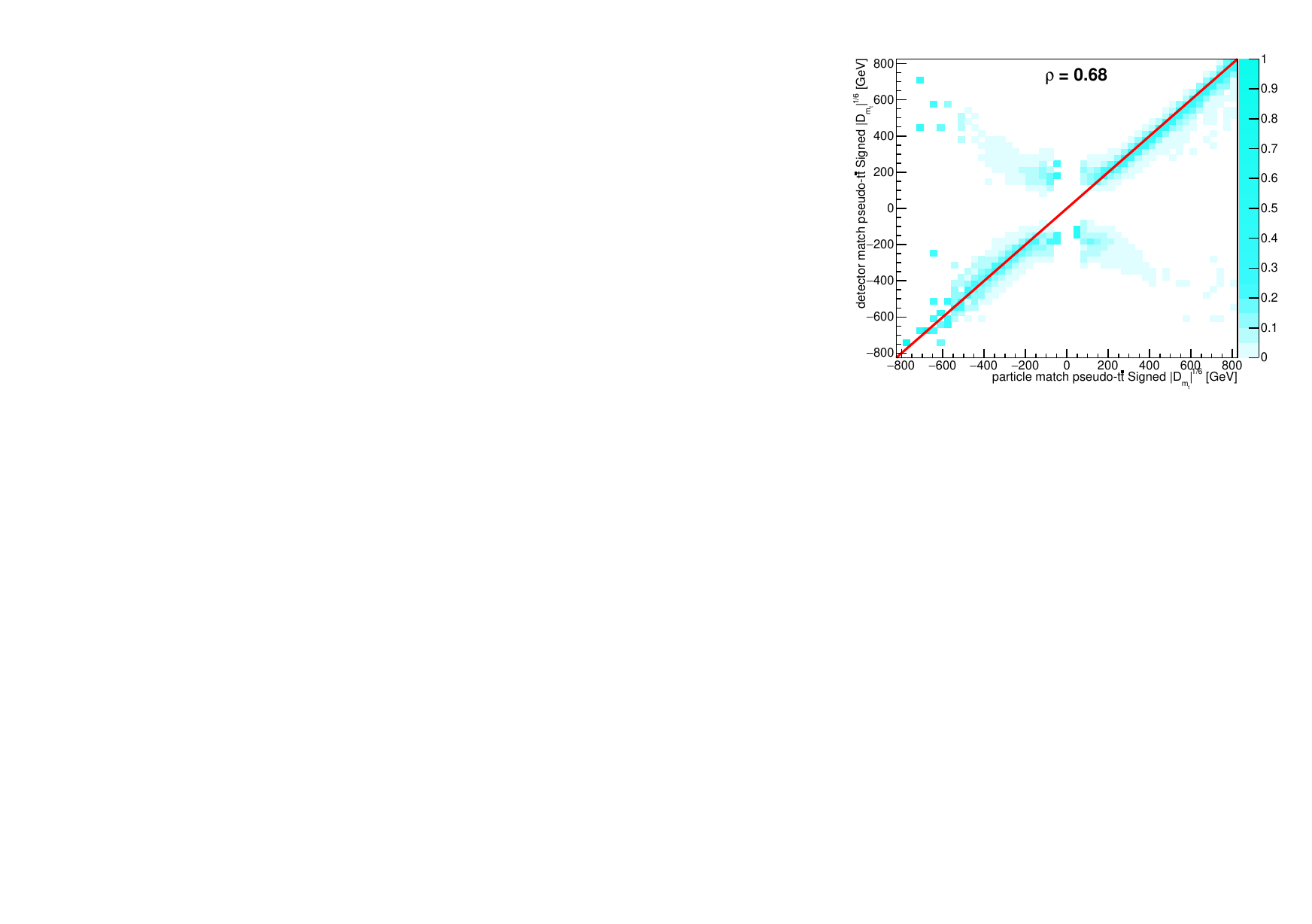} 
\caption{The migration matrix between the particle and the \Delphes{} detector levels for the signed signed discriminant (to the power of $1/6$ to keep the GeV unit) for matched events for the neutrino $p_z$ solution based on the $m_{\ell\nu}= m_W$ condition (left) and based on the $m_{t,\mathrm{had}} = m_{t,\mathrm{lep}}$ condition (right). }
\label{pst:pseudotop_Disc_study_match}
\end{figure}

 The migration matrices between the particle and detector levels (without the matching requirement) for the signed discriminant of the above quadratic equations (to the power of $1/6$ to keep the unit of GeV) are shown in Fig.~\ref{pst:pseudotop_Disc_study1} and for particle-to-detector matched events (in terms of objects forming the pseudo-top quarks, as described in Section~\ref{sec:migra}) in Fig.~\ref{pst:pseudotop_Disc_study_match}.
More detailed studies do not show large differences in the correlation between observables at the particle and detector levels when split into categories where the signs of the discriminants are the same or opposite at the two levels.
Thus, the requirement of the diagonality of discriminants cannot substitute the performance of the matching correction.

\bibliography{main}{}

\end{document}